\documentclass[%
aps,reprint,pre,
superscriptaddress,
amsmath,
amssymb,
floatfix,
]{revtex4-2}

\usepackage[english]{babel}
\usepackage{graphicx}
\graphicspath{{./figures/}}
\usepackage{dcolumn}
\usepackage{bm}

\usepackage{xcolor}
\definecolor{lightblue}{rgb}{0,0.58,0.71}
\definecolor{deepblue}{rgb}{0.0,0.,0.5}

\usepackage{mathtools}

\DeclareMathOperator{\eig}{eig}
\DeclareMathOperator{\erfi}{erfi}
\DeclareMathOperator{\erf}{erf}

\newcommand{\mrm}[1]{\mathrm{#1}}
\newcommand{\bbm}[1]{\boldsymbol{\mrm{#1}}}

\newcommand{\pp}[2]{ \frac{\partial #1}{\partial #2}}
\newcommand{\un}[0]{ \hat{\bbm{n}}}

\newcommand{\nba}[0]{\mrm{N_{Ba}}}
\newcommand{\nbr}[0]{\mrm{N_{Br}}}

\makeatletter
\newcommand*{\balancecolsandclearpage}{%
  \close@column@grid
  \cleardoublepage
  \twocolumngrid
}
\makeatother

\usepackage[]{hyperref}

\begin{document}

\title{Delayed Hopf bifurcation and control of a ferrofluid interface \\ via a time-dependent magnetic field}

\author{Zongxin Yu}
\email{yu754@purdue.edu}
\affiliation{School of Mechanical Engineering, Purdue University, West Lafayette, Indiana 47907, USA}

\author{Ivan C.\ Christov}
\email{christov@purdue.edu}
\affiliation{School of Mechanical Engineering, Purdue University, West Lafayette, Indiana 47907, USA}
\affiliation{Department of Computer Science, University of Nicosia, 46 Makedonitissas Avenue, CY-2417, Nicosia, Cyprus}

\date{\today}

\begin{abstract}
A ferrofluid droplet confined in a Hele-Shaw cell can be deformed into a stably spinning ``gear,'' using crossed magnetic fields. Previously, fully nonlinear simulation revealed that the spinning gear emerges as a stable traveling wave along the droplet's interface bifurcates from the trivial (equilibrium) shape. In this work, a center manifold reduction is applied to show the geometrical equivalence between a two-harmonic-mode coupled system of ordinary differential equations arising from a weakly nonlinear analysis of the interface shape and a Hopf bifurcation. The rotating complex amplitude of the fundamental mode saturates to a limit circle as the periodic traveling wave solution is obtained. An amplitude equation is derived from a multiple-time-scale expansion as a reduced model of the dynamics. Then, inspired by the well-known delay behavior of time-dependent Hopf bifurcations, we design a slowly time-varying magnetic field such that the timing and emergence of the interfacial traveling wave can be controlled. The proposed theory allows us to determine the time-dependent saturated state resulting from the dynamic bifurcation and delayed onset of instability. The amplitude equation also reveals hysteresis-like behavior upon time reversal of the magnetic field. The state obtained upon time reversal differs from the state obtained during the initial (forward-time) period, yet it can still be predicted by the proposed reduced-order theory. 
\end{abstract}

\maketitle

\section{Introduction}

Ferrofluids are stable colloidal suspensions of nanometer-sized magnetic particles dispersed in a nonmagnetic carrier fluid \cite{S74,R87}. The rheological behavior of these ``smart'' fluids is typically Newtonian, yet ferrofluids can flow in response to external magnetic fields \cite{R13_Ferrohydrodynamics,BCM10}. The most visually striking example of such a ``remote control'' of the fluid is the motion of the interface between a ferrofluid and air \cite{HM20}. This behavior allows for convenient, \emph{non-invasive} manipulation of ferrofluids interfaces and flows, which has motivated a number of potential applications ranging from drug delivery \cite{VFM02} to mechanical characterization of tissues \cite{SMRK17} and soft robotics~\cite{FDKXS20,AIBBM20,YLW20}.

Ferrofluids' interfacial dynamics are also widely studied from the fundamental point of view. One canonical system is a two-dimensional free surface flow confined to a Hele-Shaw cell (\textit{i.e.}, the small gap between two large, rigid plates \cite{HS98}), which provides a fertile ground for exploring nonlinear physics \cite{BKLST86}. In this context, driven ferrofluids exhibit \emph{pattern formation}. One remarkable type of pattern is the so-called labyrinthine instability \cite{RZS83,LGJ92}, caused by imposing a uniform magnetic field perpendicular to a horizontal Hele-Shaw cell. Another pattern-forming phenomenon studied analytically \cite{OML08,LM16} is a ferrofluid droplet in a Hele-Shaw cell subject to a radial magnetic field. The droplet interface experiences linear instability and evolves into a stationary starfish-like pattern. Statics and dynamics of a ferrofluid droplet in both rotating \cite{LMO10} and motionless \cite{DM15} Hele-Shaw cells subjected to an azimuthal magnetic field have been studied using weakly nonlinear analysis. Next,  to influence the interfacial mode selection, Jackson and Miranda~\cite{JM07} introduced a model ``crossed'' magnetic field, which has both perpendicular and tangential components along a free ferrofluid interface. Recently, we investigated one such magnetic field setup, showing that the crossed field (with a combination of radial and azimuthal components) leads to the ferrofluid droplet achieving a stable profile shape that further rotates with a predictable angular velocity \cite{YC21}. This configuration was further studied in the context of the unstable evolution of the droplet \cite{OCAM21}, modified for a ferrofluid annulus \cite{LAM22,CM22}, and also considered in the context of wave propagation under a thin-film long-wave equation \cite{YC21_rspa}. Despite previous work identifying the steady and periodic interfacial waves on a ferrofluid droplet under a combined radial and azimuthal magnetic field, this model problem has not been thoroughly investigated from a dynamical systems perspective.

A striking feature of pattern formation in confined ferrofluids, especially near the critical point of linear instability, is the apparent low dimensionality of the dynamics. This observation allows for a description of the complex dynamics of the fluid flow (in principle, infinite-dimensional) as a finite-dimensional system of ordinary differential equations (ODEs). The system of ODEs easily reveals the stable and unstable invariant objects in the phase space, such as steady states and periodic orbits. Canonical examples of such reductions can be traced back to the low-dimensional models of turbulence by \citet{hopf1948} and of atmospheric convection by \citet{lorenz1963}. In a Hele-Shew cell, complex behaviors of bubble evolution including symmetry breaking, bistability, and non-trivial transients, were reported by \citet{FTHJ18}. These dynamics were subsequently investigated theoretically by \citet{KTLJH19}, using a weakly nonlinear analysis in the physical domain, finding that unstable periodic orbits are edge states. Weakly nonlinear analysis can also be applied in the Fourier domain. Such a perturbative, second-order mode-coupling analysis was employed to study the pattern-forming dynamics in a Hele-Shaw cell with fluid injection \cite{MW98}, and then followed by extensive analytical studies of different control strategies (see, \textit{e.g.}, \cite{MO04,LM16,ALM18}). The Fourier-domain weakly nonlinear approach was used to identify the stationary shape~\cite{LM16} and traveling-wave profile~\cite{YC21} of a ferrofluid interface in a Hele-Shaw cell from a finite-dimensional system of ODEs, which is more computationally efficient than solving the Hele-Shaw equations along with the nonlinear interfacial conditions. However, a complete characterization of the dynamics (\textit{i.e.} the stability of the orbits, and the type of the bifurcation) is lacking for these coherent structures.

External forcing strategies, for instance, the manipulation of the rigid geometry of the Hele-Shaw cell \cite{ATS12}, using elastic-walled cells \cite{PPIHJ12}, and imposing an electric \cite{MB17} or magnetic \cite{M00} field, are effective strategies for passive control. Recently, ``nonstandard'' time-dependent control strategies are also attracting attention \cite{MMM19}. Early theoretical and experimental work by \citet{CW95} showed that the interfacial instabilities are suppressed if the injection rate in a radial Hele-Shaw flow follows a power law in time. Their idea was refined by \citet{LLFP09}, whose numerical and experimental study manipulated fingering patterns by controlling the injection rate of the less viscous fluid. 
More recently, \citet{zks2015} proposed a time-dependent strategy for manipulating the fingering pattern (instability can either be suppressed or a fingering pattern, with a prescribed number of fingers, can be selected and maintained) using a time-varying gap thickness in a lifting Hele-Shaw cell (see also \cite{STW97}). Meanwhile, \citet{azll22} designed control protocols to produce self-similar patterns in electro-osmotic flow by adjusting both the electric current and the flow rate. Similarly, time-varying external forcing is easy to achieve for ferrofluids, without altering the cell geometry. For example, \citet{JGC94} proposed a simple model using a linearly increasing magnetic field strength to achieve pattern selection.

A universal feature of time-dependent nonlinear dynamical systems is the phenomenon of \emph{bifurcation delay}. Examples include the Eckhaus instability of a stretching spatially periodic pattern~\cite{KK15,GK19} and the time-dependent dissipative Swift--Hohenberg model for crown formation during the splashing of a drop onto a liquid film~\cite{KK2014}. Finite-time evolution of a dynamic instability is characterized by two instability onset times: (i) the time at which the equilibrium loses its stability, and then (ii) the time at which the solution is repelled from the equilibrium. The nonzero difference between these two times is termed the bifurcation delay. Clearly, such a phenomenon is expected to occur for ferrofluid interfaces under time-dependent magnetic fields. However, it has not been discussed previously.

Thus, motivated by the prior studies and the knowledge gap in understanding the nonlinear dynamics and bifurcations of confined ferrofluid interfaces under time-dependent forcings, in this work, we first use a two-harmonic-mode coupled ODE system to approximate the weakly nonlinear dynamics (Secs.~\ref{sec:governing} and \ref{sec:traveling_wave}). Then, we adopt a center manifold reduction to show the geometrical equivalence between this two-mode ODE system and the Hopf bifurcation (Sec.~\ref{sec:Hopf}). Inspired by the delayed Hopf bifurcation~\cite{L91,KK15}, in which the dynamics is infinitesimally slow until a critical time, at which the system abruptly begins to oscillate with a large amplitude, we show that such time-accumulated instability can be used to manipulate pattern evolution in our ferrofluid Hele-Shaw model (Secs.~\ref{sec:multiple-time-scale} and \ref{sec:time-dependent}). Finally, conclusions are stated in Sec.~\ref{sec:conclusion}. Appendices~\ref{sec:ai bi ci di}--\ref{sec:protocols} provide further technical details and examples for the reader's convenience.

\section{Problem formulation and governing equations}
\label{sec:governing}
\begin{figure}
 \centering \includegraphics[keepaspectratio=true,width=\columnwidth]{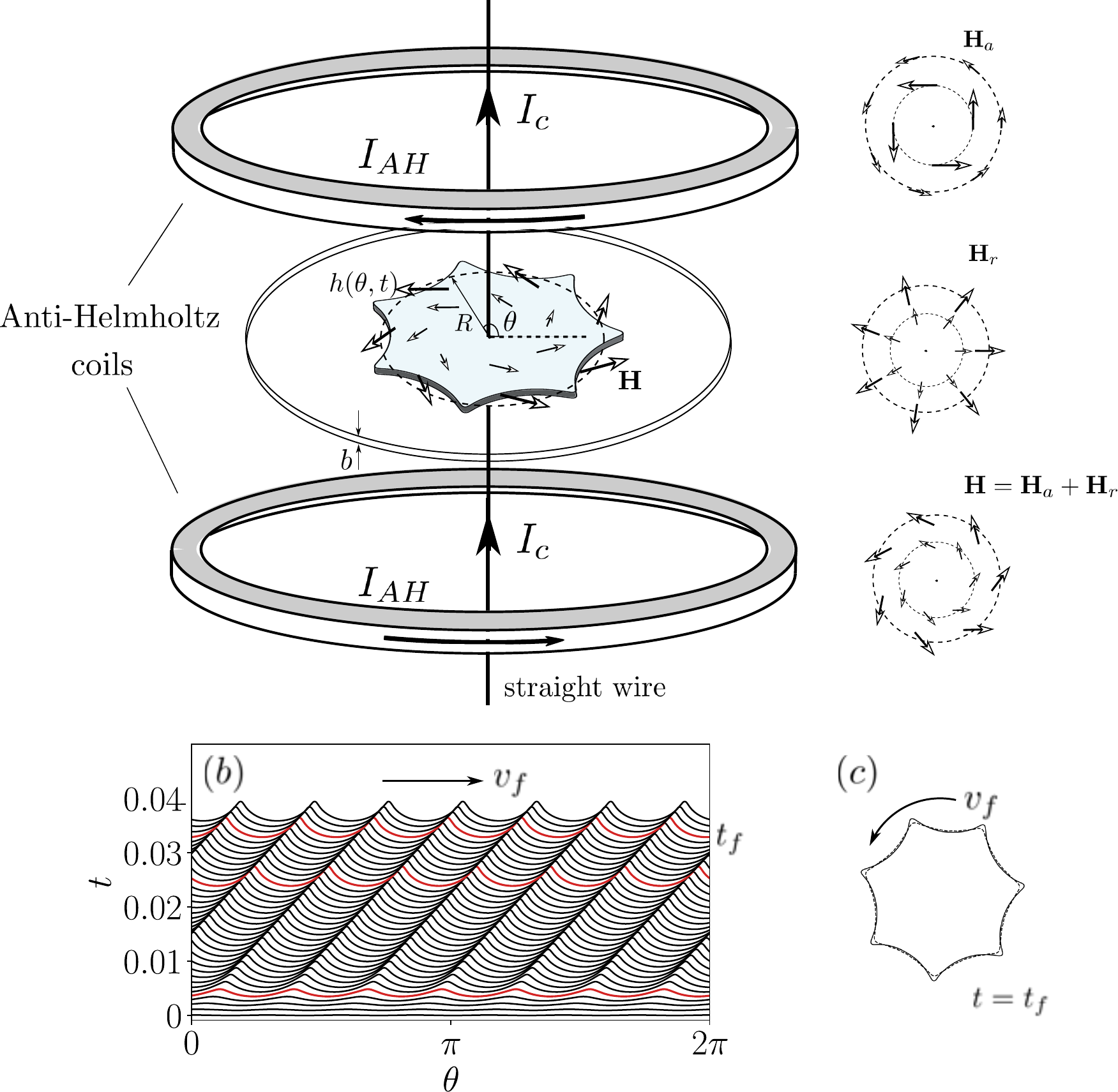}
 \caption{(\textit{a}) Schematic illustration of a horizontal Hele-Shaw cell confining a ferrofluid droplet, which is initially circular with a radius $R$. An azimuthal magnetic field $\bbm H_{a}$ is produced by a long wire conveying an electric current $I_c$. A radial magnetic field $\bbm H_{r}$ is produced by a pair of anti-Helmholtz coils with equal currents $ I_{AH}$ in opposite directions. The combined field $\bbm H = \bbm H_{a} + \bbm H_{r}$ deforms the droplet. The droplet's interface shape is given by $h(\theta,t)$. The fluid exterior to the droplet is assumed to have negligible viscosity and velocity (\textit{e.g.}, it can be taken to be air). (\textit{b}) The nonlinear evolution of the interface from a small perturbation of the flat base state ($h=R$) into a permanent traveling wave. (\textit{c}) The interfacial traveling wave causes the droplet to rotate with speed $v_f$. The motion of the droplet is sufficiently slow to neglect flow inertia. Panel images adapted, with permission, from \cite{YC21}.}
 \label{fig:config}
\end{figure}

{Inspired by an early, Cartesian model of interfacial waves driven by a ``tilted'' magnetic field \cite{LM12}, in} our previous work \cite{YC21}, we proposed a static nonuniform magnetic field configuration $\bbm H$, under which a ferrofluid droplet can deform, driven by interfacial waves, into a spinning gear. The droplet is confined in the Hele-Shaw cell, and $\bbm H$ consists of the combination of an azimuthal field and a radial field. Thus, $\bbm H$ forms an angle with the initially undisturbed (circular) interface of the confined droplet, as shown in Fig.~\ref{fig:config}(\textit{a}). {A time-dependent field can be generated by varying the magnitude and the direction of the currents in the central wire (for the azimuthal field) or in the anti-Helmholtz coils (for the radial field).} A linear closure for the ferrofluid's magnetization $\bbm M$ is usually assumed under a static or quasi-static field since the time scale of the magnetic relaxation is several orders smaller than the flow scale~\cite{JGC94,QACF18}. Thus, when we discuss the dynamics under a time-dependent field in Sec.~\ref{sec:time-dependent}, it is still under the linear magnetization assumption: $\bbm M(t) \parallel \bbm H(t)$. In the configuration shown in Fig.~\ref{fig:config}(\textit{a}), the droplet experiences both a magnetic body force and a surface traction $\propto (\bbm M\cdot \un)^2$ \cite{R13_Ferrohydrodynamics}, where $\un$ denotes the outward unit normal vector at the mobile interface. The projection of $\bbm M$ onto $\un$ breaks the symmetry of the initial droplet interface, and causes the droplet to rotate \cite{YC21}.

When linearly unstable, small perturbations of the droplet's shape grow exponentially and, then, saturate to a permanent traveling wave (causing the droplet to rotate) as shown in Fig.~\ref{fig:config}(\textit{b},\textit{c}). In \cite{YC21}, the nonlinear evolution was studied mainly through fully nonlinear simulation. The low-dimensional ODEs{, such as Eqs.~\eqref{eq:wnl} and \eqref{eq:4th order ode} to be discussed below,} arising from a weakly nonlinear analysis can also serve as a good approximation of the shape, but do not provide {dynamical} intuition beyond the initial, linear growth regime. {In this study, we first derive a simpler model, using weakly nonlinear analysis, which allows us to gain dynamical insights.}
Then, we compare this new model with the nonlinear simulations performed using a vortex-sheet solver. The vortex-sheet method is a standard sharp-interface technique for simulating the dynamics of Hele-Shaw flows \cite{R83,TA83,STW97}. It is based on a boundary integral formulation in which the fluid--fluid interface is formally replaced by a generalized vortex sheet \cite{Prosp02}. For the present problem, this type of solver was introduced and benchmarked in \cite{YC21}.

To start, we consider an initially circular interface with radius $R$ whose shape, defined as $r=h$ in the plane, is perturbed as $h(\theta,t) = R+\xi(\theta,t)$, with $\theta \in [0,2\pi]$. The perturbation $\xi$ can be expanded into Fourier modes as
\begin{equation}
    \xi(\theta,t)=\sum_{k=-\infty}^{+\infty}\xi_{k}(t)e^{ik\theta},
    \label{eq:xi_def}
\end{equation}
where $\xi_k(t)\in\mathbb{C}$ are the complex Fourier amplitudes with azimuthal wavenumbers $k\in\mathbb{Z}$. Through a weakly nonlinear analysis \cite{MW98}, the dimensionless evolution equations of the mode amplitudes, up to second order in $\xi$, can be found \cite{YC21} to be:
\begin{multline}
\dot{\xi}_k=\Lambda(k)\xi_k \\
+\sum_{k' \neq 0}  F(k,k')\xi_{k'} \xi_{k-k'} 
+ G(k,k')\dot{\xi}_{k'} \xi_{k-k'},
\label{eq:wnl}
\end{multline}
where we have defined the linear growth rate of mode $k$ as
\begin{multline}
\Lambda(k) =\frac{|k|}{R^3}(1-k^2) - \frac{2\nba }{R^4}|k|
+2(1+\chi)\nbr|k|\\
-\frac{2\chi \sqrt{\nba \nbr}}{R^2}i k |k|.
\label{eq:Linear growth}
\end{multline}
Here, $\nba$ and $\nbr$ are magnetic Bond (dimensionless) numbers that represent the ratio of the strengths of the corresponding magnetic body forces arising from the  azimuthal and radial magnetic field components, respectively, to the capillary force. The nonlinear interaction functions $F(k,k')$ and $G(k,k')$ in Eq.~\eqref{eq:wnl}, which also depend on $\nba$ and $\nbr$, are given in \cite{YC21}. {Note that under the static magnetic field in this section, $\nba$ and $\nbr$ are  constants. When the magnetic field is made time-dependent (to be discussed in Sec.~\ref{sec:time-dependent}), $\nba=\nba(t)$, $\nbr=\nbr(t)$, and $\Lambda=\Lambda(k,t)$  accordingly.}

The simulations in \cite{YC21} showed that the droplet shape exhibits a long-wave instability, and a finite number of harmonic modes can appropriately describe the dynamics. In this study, we are interested in the dynamics around the critical point, \textit{i.e.}, when the system achieves $\Re[\Lambda(k_f)]=0$, where $k_f$ is the fundamental mode (we set $k_f=7$ as in \cite{YC21}). When the fundamental mode is marginally unstable, \textit{i.e.}, $\Re[\Lambda(k_f)]\gtrsim 0$, a small number of harmonic modes is sufficient to approximate the fully nonlinear dynamics. Thus, we first truncate Eq.~\eqref{eq:wnl} with four harmonic modes, $k=k_f, 2k_f, 3k_f, 4k_f$, representing the interactions with the fundamental mode. The representation using only four harmonic modes is sufficient for the parameters used in this study. This fact will be demonstrated \textit{a posteriori} by comparison to the fully nonlinear simulation in Figs.~\ref{fig:landau}, \ref{fig:delayed prediction}, \ref{fig:stop control}, and \ref{fig:Hysteresis}.

To obtain an explicit-in-time system of equations for $\xi_k$, we further eliminate $\dot{\xi}_{k'}$ on the right-hand side of Eq.~\eqref{eq:wnl} by reusing the equation itself. We thus obtain a system of four nonlinear ODEs:
\begin{subequations}\begin{align}
\dot{x}&=a_1x +a_2 x^*y + a_3 y^*z + a_4z^*p,\\
\dot{y}&=b_1y +b_2x^*z + b_3 y^*p + b_4x^2,\\
\dot{z}&=c_1z +c_2x^*p + c_3 xy, \\
\dot{p}&=d_1p +d_2xz + d_3 y^2,
\end{align}
\label{eq:4th order ode}\end{subequations}
where $x=\xi_{k_f}$, $y=\xi_{2k_f}$, $z=\xi_{3k_f}$, $p=\xi_{4k_f}$. The superscript $*$ denotes complex conjugation. The system \eqref{eq:4th order ode} retains all second-order terms in the perturbation's amplitude. The expressions for the complex coefficients $a_j$, $b_j$, $c_j$, and $d_j$ are given in Appendix~\ref{sec:ai bi ci di}.

\section{Traveling wave solution and its stability}
\label{sec:traveling_wave}
\begin{figure}
    \centering
    \includegraphics[keepaspectratio=true,width=0.99\columnwidth]{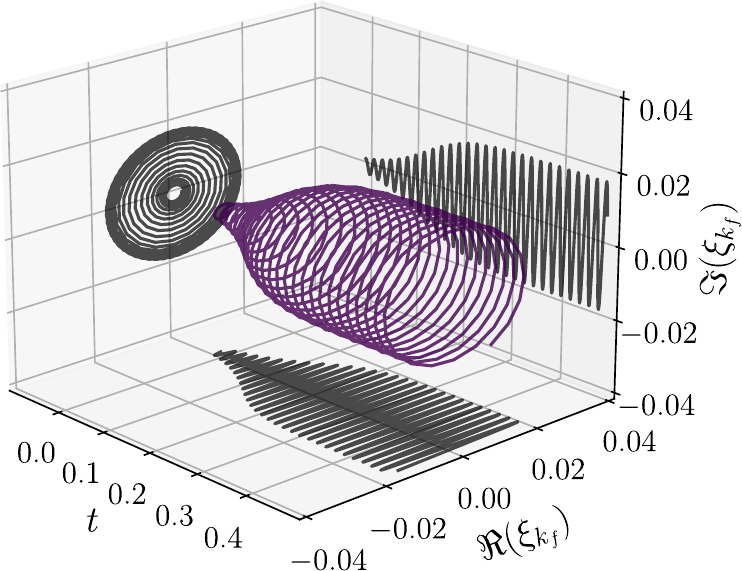}
    \caption{The evolution of the fundamental mode $k_f$ froma  fully nonlinear simulation with $\nba=1$ and $\nbr=13$.
    }
    \label{fig:energy 3D}
\end{figure}
The system \eqref{eq:4th order ode} can be  conveniently written in polar form by setting  
$j=r_j(t)e^{i\phi_j(t)}$, where $j\in\{x,y,z,p\}$.
Under this transformation, the evolution equations for the amplitudes $r_j\in\mathbb{R}$ and phase angles $\phi_j\in\mathbb{R}$ of the first four harmonic modes become decoupled, yielding separate ODEs for the real and imaginary parts of the complex ODEs. The complex ODEs are written out in Appendix~\ref{sec:polar coordinate}.

For the original droplet problem, the traveling wave solution on the periodic domain $[0,2\pi]$ can be written as $\xi(\theta,t)=\sum_{k=-\infty}^{+\infty} r_k e^{i\phi_k(t)}$, where the real amplitudes $r_k$ are independent of time and related to the complex amplitudes in Eq.~\eqref{eq:xi_def} via $\xi_k=r_ke^{-ikv_pt+\phi_{0,k}}$. The phase depends on time as $\phi_k(t)=k(\theta-v_p t)+\phi_{0,k}$, such that $\dot{\phi}_k=-kv_p$ with $v_p$ being the (right) propagation speed of the traveling wave. Here, the $\phi_{0,k}$ describe the relative phase difference with respect to the fundamental mode. One example of a fully nonlinear simulation is shown in Fig.~\ref{fig:energy 3D}, where the magnitude $r_{k_f}=|\xi_{k_f}(t)|$ of the fundamental mode's rotating complex amplitude $\xi_{k_f}(t)$ saturates to a constant as the traveling wave solution is achieved.

To understand this traveling wave solution, we set $\dot{r}_x=\dot{r}_y=\dot{r}_z=\dot{r}_p=0$ and 
\begin{equation}
\begin{alignedat}{2}
&\phi_x=\Omega t,\quad
&&\phi_y=2\Omega t +\phi_{0,y},\qquad\\
&\phi_z=3\Omega t+\phi_{0,z},\qquad
&&\phi_p=4\Omega t+\phi_{0,p},
\end{alignedat}
\label{eq:4th phase change rate}
\end{equation}
where $\Omega=-k_fv_p$ is the rate of change of the phase of the fundamental mode $k_f$. Substituting the traveling wave solution~\eqref{eq:4th phase change rate} into the system~\eqref{eq:4th order ode} (or in the polar form system~\eqref{eq:4th order ode_polar}) gives rise to:%
\begin{subequations}\begin{align}
(i\Omega-a_1)r_x 
&=a_2r_xr_ye^{iD}
+a_3r_yr_ze^{iA}
+a_4r_zr_pe^{iB},\\
(i2\Omega-b_1)r_y 
&=b_2r_xr_ze^{iA}
+b_3r_yr_pe^{iC}
+b_4r_x^2e^{-iD},\label{eq:4th stationary_2}\\
(i3\Omega-c_1)r_z  
&=c_2r_xr_pe^{iB}
+c_3r_xr_ye^{-iA},\\
(i4\Omega-d_1)r_p 
&=d_2r_xr_ze^{-iB}
+d_3r_y^2e^{-iC},
\end{align}\label{eq:4th stationary}\end{subequations}
where $A=\phi_{0,z}-\phi_{0,y}$, $B=\phi_{0,p}-\phi_{0,z}$, $C=\phi_{0,p}-2\phi_{0,y}$ are the relative phase difference. The latter three unknowns, together with $r_x$, $r_y$, $r_z$, $r_p$, and $\Omega$, characterize the nonlinear traveling wave; note that $D$ is calculated from $A$, $B$, and $C$ as $D=A+B-C$.

\begin{figure}
    \includegraphics[keepaspectratio=true,width=\columnwidth]{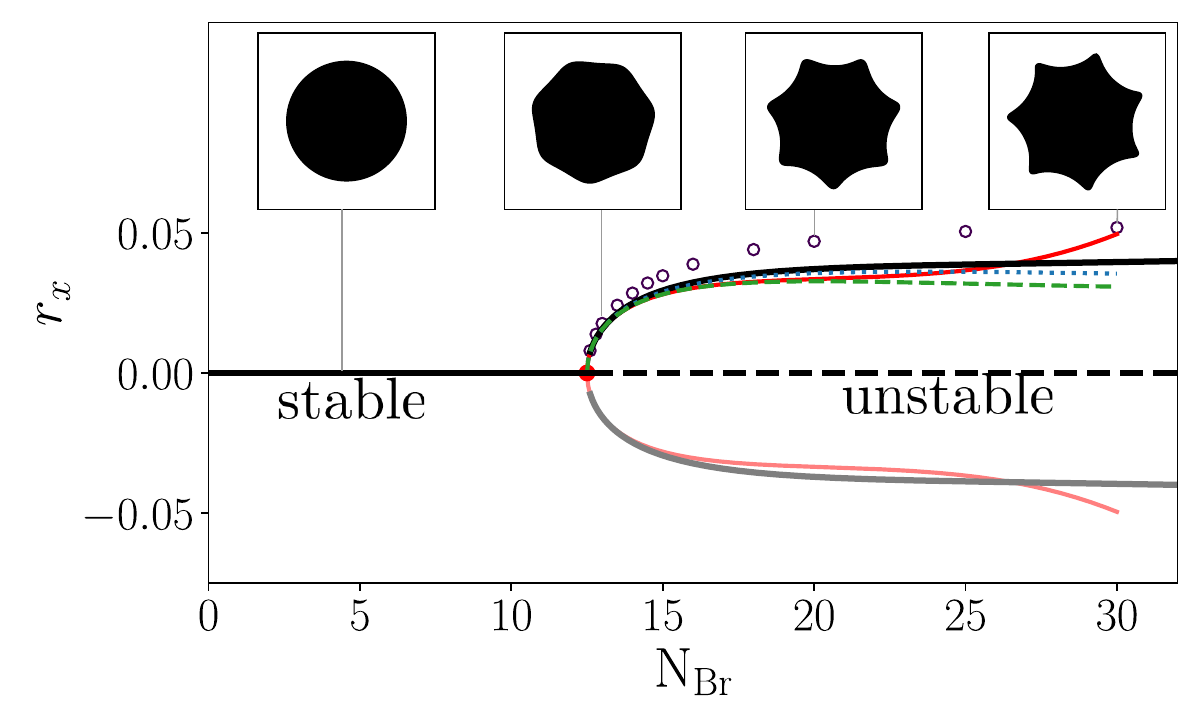}       
    \caption{The fundamental mode's amplitude bifurcates with $\nbr$. The circles mark the amplitude from the fully nonlinear simulations.
    The black and gray solid curves show the solution of the four-mode coupling system~\eqref{eq:4th stationary}, while the gray curve (with negative amplitude) has no physical meaning. The black dashed line represents the unstable trivial solution $r_j=0$, $j\in\{x,y,z,p\}$. 
    The red curve shows the solution near the critical point obtained from the reduced model~\eqref{eq:2nd stationary}. 
    The green dashed line shows the result from the center manifold reduction~\eqref{eq:center manifold}. The blue dotted line shows the multiple-time-scale analysis result from Eq.~\eqref{eq:Landau}. 
    An azimuthal field with $\nba=1$ is used, and to set the  critical point of the system, \textit{i.e.}, $\Re(a_1)=0$ for $k_f=7$, we must take $\nbr=12.5$.} 
    \label{fig:rx bifurcation}
\end{figure}

Equations~\eqref{eq:4th stationary} are solved using a Newton--Krylov method available in the SciPy library \cite{SciPy}. The solutions are shown in Fig.~\ref{fig:rx bifurcation}. Near the critical point of the system, when $\Re(a_1)=0$, the magnitudes of the higher-order modes (\textit{i.e.}, $r_z$ and $r_p$) become small (comparable to machine precision), and the Newton--Krylov method struggles to converge. 
 
On the other hand, as the nonlinearity becomes weaker, the system can be approximated by an even lower-order system. Taking $z=p=0$, the system \eqref{eq:4th order ode} reduces to 
\begin{subequations}\begin{align}
\dot{x}&=a_1x+a_2x^*y,\\
\dot{y}&=b_1y+b_4x^{2},
\end{align} \label{eq:2nd order ode}
\end{subequations}
and the stationary solution $r_x$ is found from \eqref{eq:2nd order ode} to satisfy 
\begin{equation}
    (b_1-2i\Omega)(a_1-i\Omega)=a_2b_4r_x^2.
    \label{eq:2nd stationary}
\end{equation}
This stationary solution is shown in Fig.~\ref{fig:rx bifurcation}.
One immediate conclusion that can be drawn from Eq.~\eqref{eq:2nd stationary} is that at the critical point, when $\Re{(a_1)}=0$, one solution is $\Omega=\Im{(a_1)}$ and $r_x=0$. This solution corresponds to the non-hyperbolic equilibrium point. Along this solution branch of Eq.~\eqref{eq:2nd stationary}, if $\Re{(a_1)}$ were to further decrease (and become negative), then $r_x^2<0$, and thus there are no real solutions for $r_x$. In other words, the traveling wave solution does not exist (initial perturbation to the equilibrium state decay).

Next, we address the question of the stability of the traveling wave solution. We perturb the complex stationary solution by taking 
\begin{subequations}
    \label{eq:perturbed traveling wave}
    \begin{align}
        x(t)&=\left(\epsilon_x + r_xe^{i\phi_{0,x}}\right)e^{i\Omega t}, \\
        y(t)&=\left(\epsilon_y + r_ye^{i\phi_{0,y}}\right)e^{i2\Omega t},\\
        z(t)&=\left(\epsilon_z + r_ze^{i\phi_{0,z}}\right)e^{i3\Omega t},\\
        p(t)&=\left(\epsilon_p + r_pe^{i\phi_{0,p}}\right)e^{i4\Omega t}.
    \end{align}
\end{subequations}
The evolution of the perturbation $\bm{\epsilon}=[\epsilon_x,\epsilon_y,\epsilon_z,\epsilon_p]^\top$
is given by $\dot{\bm{\epsilon}}=\bm{M}\bm{\epsilon}$, where the matrix $\bm{M}$ is given in Appendix~\ref{sec:eigenvalues}.

We find that the real part of the four eigenvalues of $\bm{M}$ is always negative for the range of parameters considered in this study (see Fig.~\ref{fig:eigenvalue} in  Appendix~\ref{sec:eigenvalues}), which indicates that the traveling wave is on the stable solution branch of the dynamical system. This result agrees with the stability diagram {numerically investigated} in \cite{YC21}, wherein the traveling wave profiles were found to be local attractors. {Also, while \cite{YC21} studied the stability of the droplet profile in the physical domain, the current study revises and verifies the result in the Fourier domain.}

The bifurcation of the amplitude $r_x$ with increasing $\nbr$ is shown in Fig.~\ref{fig:rx bifurcation}. A stable limit circle emerging from the trivial solution beyond a critical value of the parameter is, of course, the familiar Hopf bifurcation. The limit circle is the traveling wave solution with complex amplitude rotating at a constant speed $\Omega$, which is also seen in Fig.~\ref{fig:energy 3D}. Next, we wish to understand the details and implications of this Hopf bifurcation of the ferrofluid droplet's interface dynamics.

\section{Supercritical Hopf bifurcation}
\label{sec:Hopf}
The system \eqref{eq:4th order ode} of four complex-valued nonlinear ODEs is challenging to analyze. Instead, to determine the properties of the observed bifurcation, we  consider the reduced, two-mode system \eqref{eq:2nd order ode}. This reduction is supported by the fact that, around the critical point (\textit{i.e.}, for weak nonlinearity), the dynamics can be well approximated by a small number of harmonic modes. Indeed, the fully nonlinear simulation in Fig.~\ref{fig:landau}(\textit{a},\textit{b}) shows that, around the critical point (here, $\nbr=12.5$ when $\Re(a_1)=0$), the dynamics involves effectively only two harmonic modes (the fundamental mode $k=k_f=7$ and its harmonic $k=2k_f=14$). For larger $\nbr$, the ``strength'' of the instability also increases (since $a_1$ increases with $\nbr$), and nonlinearity leads to the interaction of multiple harmonics modes, as seen in Fig.~\ref{fig:landau}(\textit{c}). However, around the critical point, as in Fig.~\ref{fig:landau}(\textit{a},\textit{b}), the system~\eqref{eq:2nd order ode} captures the leading-order behavior.

The linearization of the system \eqref{eq:2nd order ode} around the fixed point $(x,y)=(0,0)$ is simply 
\begin{equation}
    \dot{x}=a_1x, \qquad \dot{y}=b_1y.
\end{equation}
Thus, the dynamics of $x$ and $y$ are decoupled. We are only interested in leading mode, for which we have:
\begin{subequations}\begin{align}
    \dot{x}_r&=\Re(a_1)x_r-\Im(a_1)x_i,\\
    \dot{x}_i&=\Re(a_1)x_i+\Im(a_1)x_r,
\end{align}\label{eq:2nd order ode_lin}\end{subequations}%
where $x_r=\Re(x)$ and  $x_i=\Im(x)$.
The linearized system~\eqref{eq:2nd order ode_lin} has a pair of eigenvalues $\lambda_\pm=\Re(a_1) \pm i\Im(a_1)$. Thus, the non-hyperbolicity condition (\textit{i.e.}, that one conjugate pair of imaginary eigenvalues exist at the critical point when $\Re(a_1)=0$ and $\Im(a_1)\neq 0$), and the transversality condition (\textit{i.e.}, that ${\partial \Re(a_1)}/{\partial \nbr}\neq 0$) of the Hopf bifurcation are easily verified. To satisfy the genericity condition, however, the first Lyapunov coefficient needs to also be shown to be negative \cite{KKK98}, such that the limit cycle is orbitally stable. However, the calculation of this coefficient is not trivial for higher-dimensional systems \cite{KKK98,WWG03}. Instead, we turn to the center manifold method to further reduce the dimensionality of the system \eqref{eq:2nd order ode} near the critical point and obtain a planar dynamical system.

\subsection{Center manifold reduction}
From the dynamics studied above, we expect the current system to have a parameter-dependent center manifold on which the system exhibits the Hopf bifurcation. In contrast, the behavior off the manifold is ``trivial" (meaning that the leading mode dominates the dynamics).

\begin{figure}
    \centering
    \includegraphics[keepaspectratio=true,width=0.99\columnwidth]{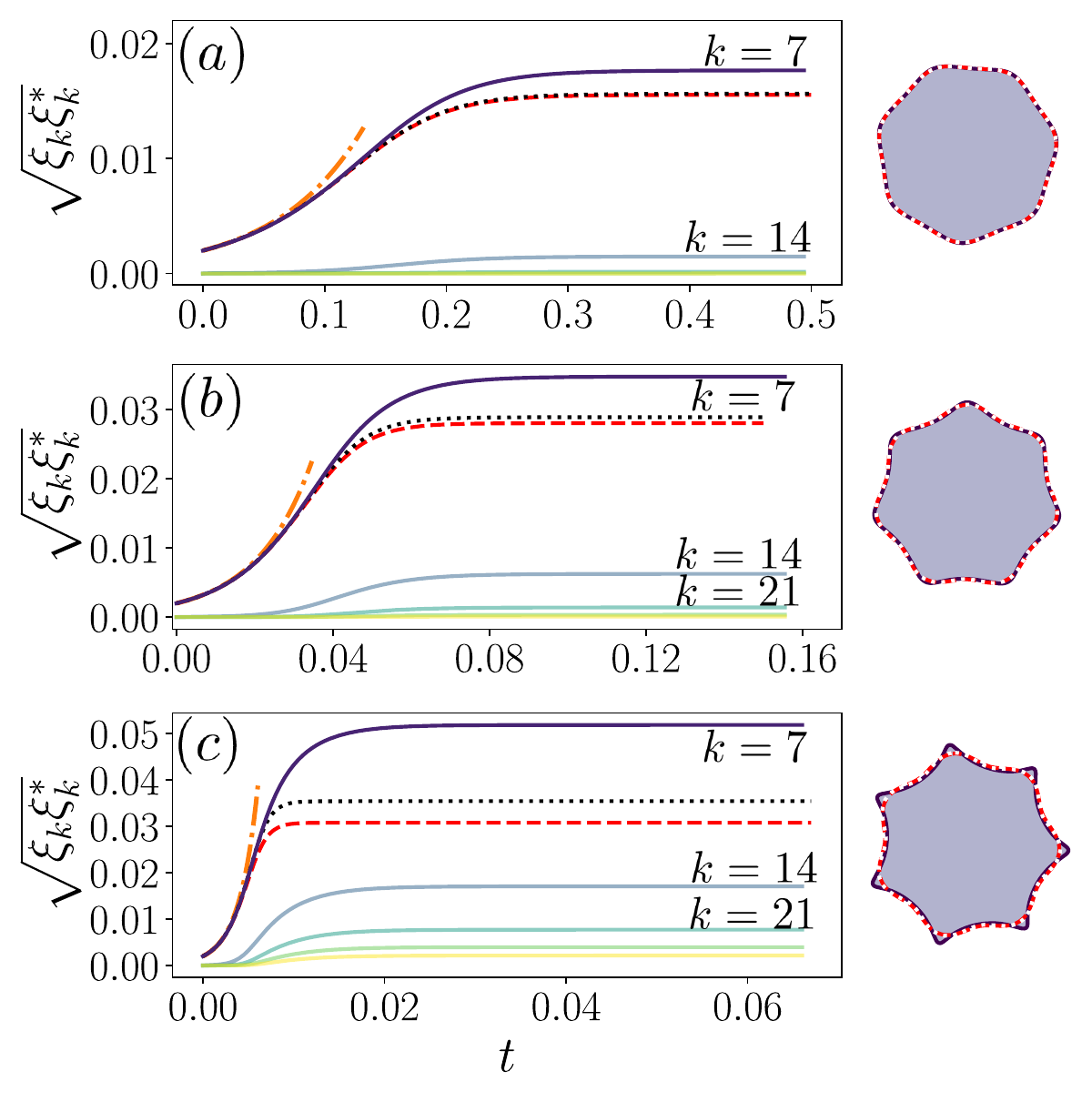}
    \caption{Comparison of leading modes' amplitude evolution for (\textit{a}) $\nbr=13$, (\textit{b}) $\nbr= 15$, and (\textit{c}) $\nbr=30$. Shown are the center manifold reduction solution from Eq.~\eqref{eq:center manifold} (black dotted curve), the multiple-time-scale analysis solution from Eq.~\eqref{eq:Landau} (red dashed curve), and the fully nonlinear simulation (solid curves). The orange dash-dotted curve shows the unstable linear evolution. The corresponding permanent rotating droplet shapes are shown on the right, produced via a fully nonlinear simulation (purple solid), via the multiple-time-scale analysis (red dashed), and via the center manifold method (white dotted).
    }
    \label{fig:landau}
\end{figure}

A quadratic approximation is used to derive the finite-dimensional center manifold \cite{KKK98,WWG03}. Specifically, we assume the dynamics on the center manifold can be related by a scalar quadratic function $y=V(x,x^*)$. For the system \eqref{eq:2nd order ode} near its critical point $(x,y)=(0,0)$, we find the center manifold (see Appendix~\ref{sec:Center manifold derivation}) to be:
\begin{equation}
    \mathcal{W}_c=\left\{(x,y):y=V(x)=\frac{b_4}{2a_1- b_1}x^2\right\}.
\end{equation}
Correspondingly, we have a locally topologically equivalent dynamical system \cite{KKK98}:
\begin{subequations}\begin{align}
    \dot{x}&=a_1x+\frac{a_2b_4}{2a_1- b_1}|x|^2x,
\label{eq:cm x}\\
     \dot{y}&=2a_1y. \label{eq:cm y}
\end{align}\label{eq:center manifold}\end{subequations}%
Now, the equations for $x$ and $y$ are decoupled and Eq.~\eqref{eq:cm x} is the restriction \cite{KKK98} of the system \eqref{eq:2nd order ode} to its center manifold $\mathcal{W}_c$. The dynamics of the system are essentially determined by this restriction, \textit{i.e.}, Eq.~\eqref{eq:cm x},
since \eqref{eq:cm y} is linear and its dynamics is trivial. 
Indeed, as shown in Fig.~\ref{fig:landau}, Eq.~\eqref{eq:cm x} accurately captures the evolution of $x$ from system~\eqref{eq:2nd order ode} along the center manifold. It is also evident that Eq.~\eqref{eq:cm x} even captures the original fully nonlinear system's dynamics (\textit{i.e.}, equations (2)--(4) from \cite{YC21}). Further, the single ODE \eqref{eq:cm x} from the center manifold reduction also accurately predicts the permanent rotating droplet profile, especially near the critical point ($\nbr=12.5$ as in Fig.~\ref{fig:landau}(\textit{a},\textit{b})).

Notably, it takes four steps of reduction to obtain the single ODE~\eqref{eq:cm x} from the original Hele-Shaw problem. First, we performed the weakly nonlinear expansion \eqref{eq:wnl} in the Fourier domain. Second, the weakly nonlinear expansion was truncated at a finite number of harmonic modes (four in the current study), to yield the system~\eqref{eq:4th order ode}. Third, we approximated the system~\eqref{eq:4th order ode} by the  two-harmonic-mode system~\eqref{eq:2nd order ode} near the critical point. Fourth, along the center manifold, the system~\eqref{eq:2nd order ode} becomes decoupled, and the leading mode's nonlinear evolution is accurately described by Eq.~\eqref{eq:cm x}. The second and third steps can be combined since they only depend on how many modes we wish to retain. In the  physical system under consideration here, for weaker nonlinearity, a smaller number of interacting modes is present. Note that the system~\eqref{eq:2nd order ode} can also be obtained by restricting the system~\eqref{eq:4th order ode} to its critical eigenspace $\{z=0,\;p=0\}$. This \emph{tangent approximation} does not always guarantee topological equivalence \cite{KKK98,WWG03}. In the present problem, the specific meaning of the harmonic amplitudes, \textit{i.e.}, $x$, $y$, $z$, $p$, and the long-wave instability feature of the Hele-Shaw problem ensure the tangent approximation is successful.

\subsection{Normal form of the Hopf bifurcation}
Let $a_1=\mu+i\omega$ ($\omega<0$) and $\tau=-\omega t$, then Eq.~\eqref{eq:cm x} can be rewritten as:
\begin{equation}
\frac{d x}{d\tau} =\left(-\frac{\mu}{\omega}-i\right)x+\frac{a_2b_4}{-(2a_1- b_1)\omega}|x|^2x,
 \label{eq:Hopf normal form}
\end{equation}
which is the normal form of a Hopf bifurcation \cite{KKK98} in which the motion along the limit cycle is counterclockwise. The rotation direction of our ferrofluid droplet is determined by the direction of the magnetic field's azimuthal component, and thus the sign of the imaginary part of the linear growth rate, as discussed in \cite{YC21}. This sign does not change the stability of the system. 
For a dynamical system in the form \eqref{eq:Hopf normal form}, the first Lyapunov coefficient can be directly computed as $\Re\left[\frac{a_2b_4}{-(2a_1- b_1)\omega}\right]$, and shown to be always negative for the parameters chosen in this study. 
Thus, together with the condition $-
\mu/\omega>0$, the existence of a supercritical Hopf bifurcation is proven. The corresponding stable limit cycle has radius
\begin{equation}
    r_x=\sqrt{\frac{-\mu}{\Re\left(\frac{a_2b_4}{2a_1- b_1}\right)}}.
\end{equation}
As expected, Fig.~\ref{fig:rx bifurcation} shows that this radius can predict the amplitude of the traveling wave solution near the critical point of the system, \textit{i.e.}, when the ferrofluid interface experiences weak nonlinearity. 

Equation~\eqref{eq:Hopf normal form} reveals that the linearly unstable but nonlinearly stable interfacial dynamics of the confined ferrofluid interface emerge via a Hopf bifurcation. We expect that this analysis can also be applied to other Hele-Shaw problems involving interfacial dynamics characterized by long-wave instability, {such as the configuration in \cite{LM12}}. For systems exhibiting a long-wave instability, a finite set of wavenumbers usually dominates the dynamics, and thus the truncation to a finite-dimensional space, in the Fourier domain, is fruitful, reducing the original infinite-dimensional partial differential equations to a low-dimensional system of ODEs. Furthermore, in the weakly nonlinear regime, the number of unstable modes can be controlled such that two-mode interaction~\eqref{eq:2nd order ode} can be analyzed via a center manifold reduction, while still revealing important dynamical features of the original infinite-dimensional problem, {which has nonlocal dynamics as already hinted by the vortex-sheet formulation of the problem \cite{TA83,YC21}.}

{The success of the center manifold reduction may appear surprising. The simple local equation~\eqref{eq:cm x} successfully captures the nonlocal dynamics. This feature can be understood by considering the stationary pattern emerging from the balance of capillary and centrifugal forces, discussed by \citet{AOC04}. For the stationary pattern, imposing the zero vorticity condition, the vortex-sheet formulation is reduced to a single geometric ODE in space. The solution of this geometric ODE is the well-known family of \textit{elasticas}. \citet{AOC04} build the connection between the elastica solutions of the Saffman--Taylor problem and the bifurcation analysis of interfacial growth problems. The unstable branch of the subcritical bifurcation diagram obtained from their amplitude equation is similar to Eq.~\eqref{eq:cm x} herein. It is interesting to note that while Ref.~\cite{AOC04} shows the linearly stable modes in the Saffman--Taylor problem are generically nonlinearly unstable (characterized by a subcritical bifurcation), the current study finds patterns that are nonlinearly stable (characterized by a supercritical bifurcation), even if linearly unstable. However, even though the vortex-sheet formulation of the problem from \cite{AOC04} and the present study are similar, a geometric ODE providing exact solutions cannot be obtained in the current work due to the dynamic nature (\textit{i.e.}, the nonzero interface velocity and local vorticity).}

Although the proposed model reduction process, starting with the leading-order weakly nonlinear approximation and followed by the center manifold calculation, looks straightforward, it does not mean that the Hopf bifurcation result follows trivially. First, a complex linear growth rate is necessary such that, near the critical point of the system, a simple pair of complex-conjugate eigenvalues cross the imaginary axis when varying the controllable bifurcation parameter. The latter ensures the satisfaction of the non-hyperbolicity and transversality conditions. {For example, when the linear growth rate is purely real (\textit{e.g.}, when the interface is subjected to only a radial magnetic field as in \cite{OML08,LM16}), a supercritical pitchfork bifurcation can be expected, from which a \emph{static} gear-like pattern emerges. In comparison, the propagating interfacial wave, driven by the tilted magnetic field introduced in  \cite{LM12}, is expected to be governed by a Hopf bifurcation.}
In addition, it must be properly shown that the physical configuration and parameters yield a negative first Lyapunov coefficient, which ensures that a stable limit cycle emerges from the bifurcation.

Another possibility is a \emph{dynamical} bifurcation, such as a \emph{delayed} bifurcation \cite{BER89,L91}. In a delayed Hopf bifurcation, the dynamics is infinitesimally slow with respect to the bifurcation parameter. The real part of the linear growth rate is initially negative until a critical time, thereupon becoming positive, which causes the solution to abruptly begin to rotate with a large amplitude. Next, we would like to understand if a delayed bifurcation can be observed in the confined ferrofluid droplet problem. Further, we would like to determine how well the critical (delay) time can be approximated. To answer these questions, we first conduct a multiple-time-scale analysis of Eq.~\eqref{eq:2nd order ode}. Then, we analyze a time-dependent problem with a slow-varying bifurcation parameter.

\section{Multiple-time-scale analysis}
\label{sec:multiple-time-scale}
Multiple-time-scale analysis allows for the calculation of the leading effect of nonlinearity on the propagation of a harmonic wave \cite{KC96}. Following the approach used in \cite{YC21_rspa}, to begin the multiple-time-scale analysis we perturb the bifurcation parameter with $a_1 = \epsilon^2\varkappa + i\omega$ around its critical value $\Re(a_1)=0$, where again $\epsilon \ll 1$ is a small perturbation parameter, and $\varkappa>0$ is independent of $\epsilon$. The assumption that the linear growth rate is much smaller than the oscillation rate, \textit{i.e.}, $\epsilon \ll 1$ is supported by Fig.~\ref{fig:energy 3D}, in which the envelope and oscillations are clearly evolving on disparate time scales. This perturbation makes the leading mode marginally unstable and also the only unstable mode of the system. We first rescale Eq.~\eqref{eq:2nd order ode} to a small amplitude problem via $x\mapsto \epsilon x$ and $y\mapsto \epsilon y$:
\begin{subequations}\begin{align}
\dot{x}&=(\epsilon^2\varkappa+i\omega)x+\epsilon a_2x^*y,\\
\dot{y}&=b_1y+\epsilon b_4x^{2}.
\end{align} \label{eq:small amplitude}\end{subequations}
Then, we assume that $x$ and $y$ have multiple-time-scale pertubation expansions in the form:
\begin{subequations}\begin{align}
  x(t,T_1) &= x_0(t,T_1)+\epsilon x_1(t,T_1) +\epsilon^2 x_2(t,T_1) +\mathcal{O}(\epsilon^3),\\
  y(t,T_1) &= y_0(t,T_1)+\epsilon y_1(t,T_1) +\epsilon^2 y_2(t,T_1) +\mathcal{O}(\epsilon^3).
\end{align}\label{eq:expansion}\end{subequations}
The slow time scale is $T_1=\epsilon^2 t$, and the time derivative transforms as $\dot{(\cdot)}={d(\cdot)}/{dt}={\partial (\cdot)}/{\partial t}+\epsilon^2 {\partial(\cdot)}{\partial T_1}$. Substituting the time derivative and the expansion~\eqref{eq:expansion} into the small amplitude equation~\eqref{eq:small amplitude} gives rise to a series of problems at each order in $\epsilon$.

The leading-order problem, at $\mathcal{O}(1)$, is
\begin{subequations}\begin{align}
 \pp{x_0}{t}-i\omega x_0&=0,\\
  \pp{y_0}{t}-b_1 y_0&=0,
\end{align}\label{eq:O(0) eq}\end{subequations}
which has a solution of the form
\begin{subequations}\begin{align}
 x_0(t,T_1)&=A_x(T_1)e^{i\omega t},\\
y_0(t,T_1)&=A_y(T_1)e^{b_1 t},
\end{align}\label{eq:O(0) sol}\end{subequations}
subjected to the initial conditions $ x_0(0,0)=A_x(0)=X$, $y_0(0,0)=A_y(0)=Y$, where $X,Y\in\mathbb{C}$.
By eliminating secular terms at $\mathcal{O}(\epsilon^2)$ {(see Appendix~\ref{sec:multi-scale} for details)}, we obtain the complex amplitude equation:
\begin{equation}
   \frac{d A_x}{dT_1} =\varkappa A_x +\frac{a_2b_4}{2i\omega-b_1} |A_x|^2A_x.
    \label{eq:complex ampl eq}
\end{equation}
The complex amplitude $A_x(T_1)$ describes the slow temporal modulation of the base periodic (harmonic wave) solution.

Let $A_x(T_1) =\alpha(T_1) e^{i\beta(T_1)}$,
where $\alpha,\beta\in\mathbb{R}$, then the real part of the amplitude equation \eqref{eq:complex ampl eq} is
\begin{equation}
\frac{d \alpha}{dT_1} = \varkappa\alpha +Q\alpha^3,
\label{eq:Landau}
\end{equation}
where we defined $Q=\Re(\frac{a_2b_4}{2i\omega-b_1})$. The amplitude equation~\eqref{eq:Landau} is also known as the \emph{Landau equation} \cite{landau1944}. Unsurprisingly, equation \eqref{eq:Landau} agrees with the center manifold reduction~\eqref{eq:cm x}. The only difference is the denominator of $Q$. In the case of Eq.~\eqref{eq:cm x}, the derivation is limited to dynamics near the critical point, \textit{i.e.}, in the neighborhood of $\Re(a_1)=0$ with $\Re(a_1)>0$ (see \cite{KKK98}), thus $a_1$ appears in the equation. Meanwhile, Eq.~\eqref{eq:Landau} is derived by separating the real part $\varkappa$ and imaginary part $\omega$ of $a_1$ into different orders of $\epsilon$, such that there is only $i\omega$ in the denominator of $Q$. 
However, this difference is trivial. As seen in Fig.~\ref{fig:rx bifurcation}, the difference between the traveling wave amplitudes computed from Eq.~\eqref{eq:Landau} and Eq.~\eqref{eq:cm x} are barely distinguishable. Importantly, Eqs.~\eqref{eq:cm x} and \eqref{eq:Landau} are asymptotically equivalent as $\Re(a_1)\rightarrow 0$ (at the critical point).

\section{Time-dependent problem}
\label{sec:time-dependent}
A central question concerning pattern formation in time-dependent systems is how unsteady external forces affect the phase space structures and their evolution. This question is somewhat analogous to the question of how the quasistatic variation of a bifurcation parameter affects local attractors. The key insight is provided by the supercritical Hopf bifurcation, for which the instability onset (when the solution is repelled from the equilibrium) occurs later than the instant when the equilibrium loses its stability.

Above, we have shown that the amplitude equation~\eqref{eq:Landau} can predict the permanent rotating shape (traveling wave profile) seen in the nonlinear simulations of the confined ferrofluid droplet. Now, we move on to the question of dynamics: using the bifurcation delay feature to dynamically control the time evolution of the interface.

To start, we first reconsider the amplitude equation~\eqref{eq:Landau} for time-varying magnetic fields. Above, we took $a_1=\epsilon^2 \varkappa +i\omega$, where $\varkappa\in\mathbb{R}$. Now, instead, consider the slowly-varying time-dependent real growth rate $\varkappa=\varkappa(T_1)=\varkappa_0+IT_1$, and thus 
\begin{equation}
 a_1=a_1(T_1)=\epsilon^2 (\varkappa_0 +I T_1)+i\omega,
\end{equation}
where $\epsilon^2 \varkappa_0$ is the small initial growth rate, which can be either positive or negative. Here, $\epsilon^2 I$ is the slow evolution rate of $\Re(a_1)$ on the long time scale $T_1$, \textit{i.e.}, $d \Re(a_1)/d T_1=\epsilon^2 I$. {Physically, the linear variation of $a_1$ with $T_1$ can be achieved by controlling the combination of azimuthal and radial magnetic field strengths, or $\nba(t)$ and $\nbr(t)$, respectively. For example, we can take $\nba=1$ and set $\nbr(t)$ to be a suitable linear function of time.}

Next, the amplitude equation~\eqref{eq:Landau} can be shown, via the same analysis as before, to take the form:
\begin{equation}
    \frac{d\alpha}{dT_1} = (\varkappa_0+IT_1)\alpha +Q\alpha^3,
\label{eq:forward landau}
\end{equation}
and its solution is given by
\begin{multline}
    \alpha(T_1)=\exp\left(\varkappa_0T_1+\tfrac{1}{2}IT_1^2\right)\\
    \times\left[\left(-Q \sqrt{\frac{\pi}{I}}
    e^{\frac{-\varkappa_0^2}{I}}\right)\erfi\left(\frac{\varkappa_0+IT_1}{\sqrt{I}}\right)+c
    \right]^{-1/2},
\label{eq:t-landau-sol}
\end{multline}
where $c=1/X^2+Q\sqrt{\pi/I}
e^{-\varkappa_0^2/I}\erfi(\varkappa_0/\sqrt{I})$ is a constant related to the initial value $X=\alpha(T_1=0)$. The imaginary error function $\erfi$ is defined via $\erfi(z)=-i \erf(iz)$ \cite{erfi}.

\begin{figure}
    \centering
    \includegraphics[keepaspectratio=true,width=0.99\columnwidth]{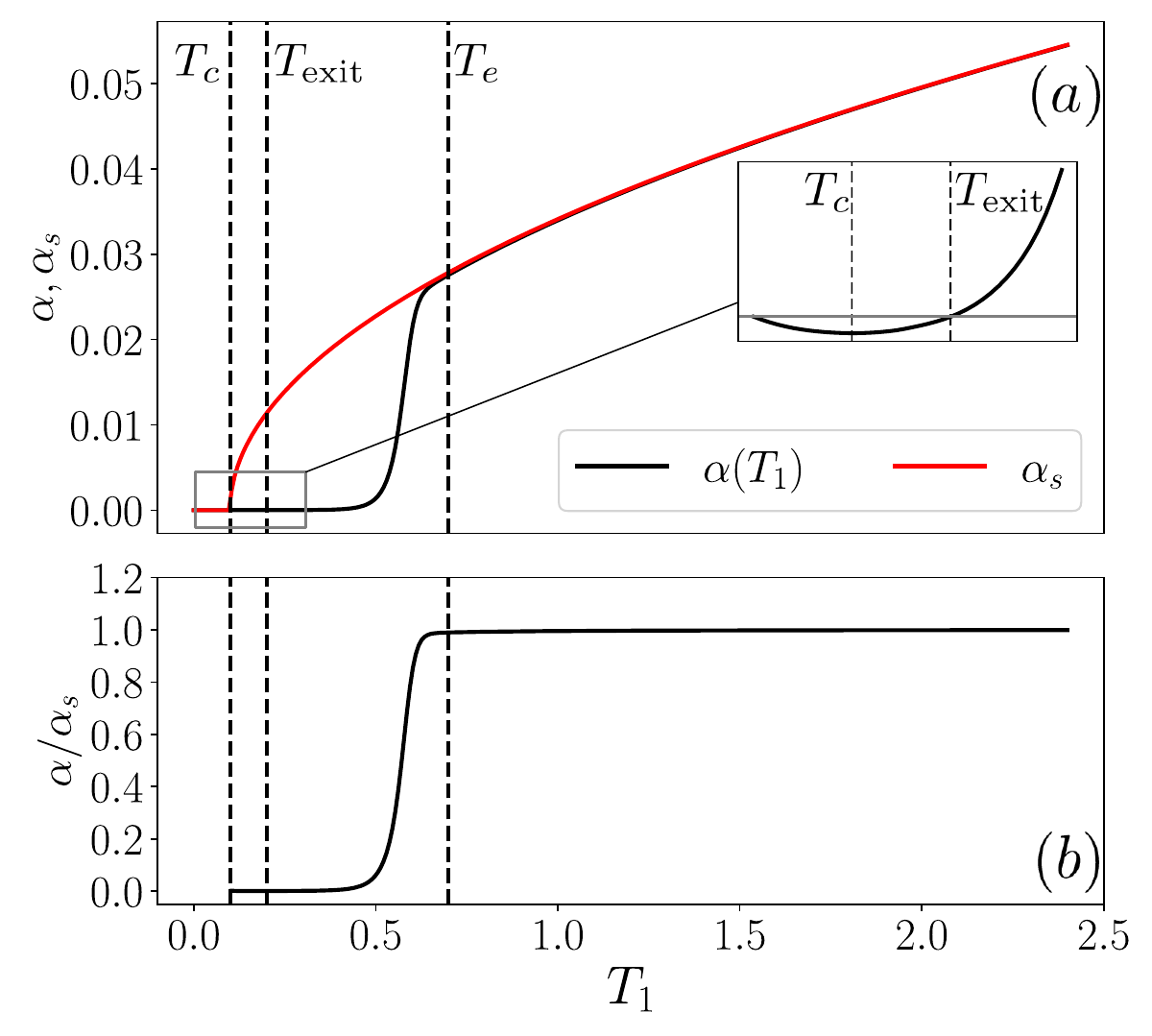}
    \caption{(\textit{a}) The solution $\alpha(T_1)$ from Eq.~\eqref{eq:t-landau-sol} (black) saturates to the quasistatic solution $\alpha_s$ from Eq.~\eqref{eq:alpha_s} (red) as $T_1$ increases. 
    (\textit{b}) The ratio $\alpha/\alpha_s$ approaches 1 during the same time period. Here, $T_{e}$ denotes the time when $\alpha/\alpha_s=\rho=0.99$. The remaining parameters are taken as $\varkappa_0=-7.5$, $I=75$, and the initial condition is $X=5\times 10^{-6}\ll1$.
    }
    \label{fig:bifur delay}
\end{figure}

The solution~\eqref{eq:t-landau-sol} for $\varkappa_0<0$ is shown in Fig.~\ref{fig:bifur delay}(\textit{a}). For $T_1<T_{c}$, the linear growth rate is such that $\Re(a_1)<0$, and the initial small perturbation decays, as shown in the inset. At $T_1=T_c$, $a_1(T_{c})=0$ and the equilibrium loses its linear stability. Now, the amplitude starts to grow, yet it remains infinitesimally small with respect to the initial perturbation. Next, at $T_1=T_{\mathrm{exit}}(>T_{c})$, the initial perturbation amplitude is recovered, and now the solution starts to repel from the initial state. Subsequently, the amplitude increases abruptly due to the positive linear growth rate. This exponential increase is also observed in the time-independent problem, as shown in Fig.~\ref{fig:landau}, which is followed by the saturation of the energy (\textit{i.e.}, emergence of the permanent traveling wave profile).

Under the proposed time-dependent field, the exponential increase is followed by a slow increase, which is identified as the quasistatic region, in which the solution slowly varies with the bifurcation parameter. As seen from Fig.~\ref{fig:bifur delay}(\textit{a}), the time-dependent solution \eqref{eq:t-landau-sol} saturates to the quasistatic solution
\begin{equation}
    \alpha_s=\sqrt{\frac{\varkappa_0+IT_1}{-Q}},
    \label{eq:alpha_s}
\end{equation}
which is obtained by setting $d\alpha/dT_1=0$ in Eq.~\eqref{eq:forward landau}. 
This saturation can be intuitively understood as the balance of the exponential factor $e^{IT_1^2/2}$ and the decay factor $\erfi[(\varkappa_0+IT_1)/\sqrt{I}]^{-1/2}$ as $T_1\rightarrow \infty$ in the time-dependent solution \eqref{eq:t-landau-sol}. This balance also provides the possibility of predicting the delay time $T_{e}$ analytically.

\subsection{Approximation of the bifurcation delay time}
To approximate the delay time $T_e$, we consider the equation
\begin{equation}
    \frac{\alpha(T_{e})}{\alpha_s(T_{e})}=\rho,
    \label{eq:alpha ratio}
\end{equation}
such that when $T_1>T_{e}$, $\alpha/\alpha_s>\rho$. In this study, we take $\rho=0.99$ without loss of generality. Now, we would like to determine $T_e$ from Eq.~\eqref{eq:alpha ratio} and establish the quality of this approximation. To this end, we use the quasistatic solution~\eqref{eq:alpha_s} and time-dependent solution~\eqref{eq:t-landau-sol} to calculate the ratio
\begin{equation}
\frac{\alpha_s^2}{\alpha^2} = 
1 + \frac{1}{2}\frac{I}{\varkappa^2} - 
\left(
\frac{1}{QX^2} e^{\varkappa_0^2/I}\right)
e^{-\varkappa^2/I}R + \mathcal{O}(\varkappa^{-4}),
\label{eq:alpha expansion}
\end{equation}
where the expansion is valid for $\varkappa=\varkappa_0+IT_1 \rightarrow \infty$. The following expansion of the imaginary error function at infinity (as $|z|\rightarrow \infty$) \cite{erfi} has been used:
\begin{equation}
    \erfi(z) = \frac{e^{z^2}}{\sqrt{\pi}}
    \left(z^{-1}+\frac{1}{2} z^{-3}+\frac{3}{4} z^{-5}+\cdots\right)-i.
\end{equation}
Further, the coefficient $\sqrt{{\pi}/{I}}\left[\erfi\left({\varkappa_0}/{\sqrt{I}}\right)+i\right]$ of the exponentially decaying term $e^{-\varkappa^2/I}\varkappa$ is neglected when compared to terms of $\mathcal{O}(1/X^2)$ for $X\ll 1$.

Note that Eq.~\eqref{eq:alpha_s} is valid only if $\varkappa_0+IT_1>0$ $\forall T_1$, \textit{i.e.}, $\varkappa_0>0$. Thus, when $\varkappa_0>0$, the time $T_{e}$ can be evaluated via Eqs.~\eqref{eq:alpha expansion} and \eqref{eq:alpha ratio}. Specifically, $T_e$ solves
\begin{equation}
    1+\frac{I}{2}\frac{1}{(\varkappa_0+IT_{e})^2}
    +\frac{\varkappa_0+IT_{e}}{-Q}\frac{1}{X^2}
    e^{-(2\varkappa_0T_{e}+IT_{e}^2)}\approx\frac{1}{\rho^2}, 
    \label{eq:T1e eval}
\end{equation}
For $\varkappa_0<0$, $T_e$ can instead be written as $T_{e}=T_{c}+T_{e,2}$, where $T_{c}=-\varkappa_0/I$ is the critical time defined by requiring a vanishing linear growth rate ($\varkappa(T_{c})=0$). When $T_1<T_{c}$, $\varkappa<0$, and perturbations decay. Thus, we can use the approximation $d \alpha/dT_1=(\varkappa_0+IT_1)\alpha$. At $T_1=T_{c}$, the initial perturbation $X$ decreases to its  minimum value of $\alpha_c$, where
\begin{equation}
   \alpha_c = X \exp\left(\int_0^{T_{c}} \varkappa_0 + IT_1 \,dT_1\right) = X e^{-\varkappa_0^2/(2I)}.
\end{equation}
For $T_1>T_{c}$, $\varkappa>0$, Eq.~\eqref{eq:T1e eval} can be used to evaluate $T_{e,2}$, by substituting $X_2=\alpha_c$ as the initial value and $\varkappa_{0,2}=0$.

\begin{figure}
    \centering
    \includegraphics[keepaspectratio=true,width=\columnwidth]{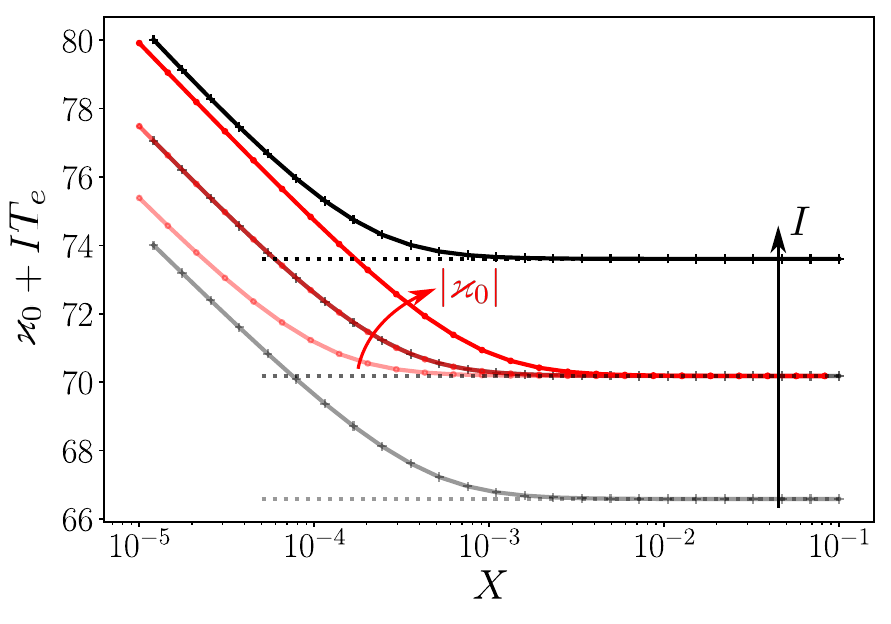}
    \caption{Dependence of $\varkappa_0+IT_{e}$ on the initial perturbation strength $X$, based on the prediction of the delay time $T_{e}$ via  Eq.~\eqref{eq:T1e eval}. The black curve with `+' markers represents the predicted time for different values of $I=180,200,220$, with an arrow pointing in the direction of increasing $I$. The red curve with `$\circ$' markers represents the predicted time for different values of $\varkappa_0=-35,-40,-45$. The dotted horizontal lines denote the asymptotic values of $\sqrt{I\rho^2/2(1-\rho^2)}$.
    } 
    \label{fig:bifur time}
\end{figure}

\begin{figure*}
    \centering
    \includegraphics[keepaspectratio=true,width=\textwidth]{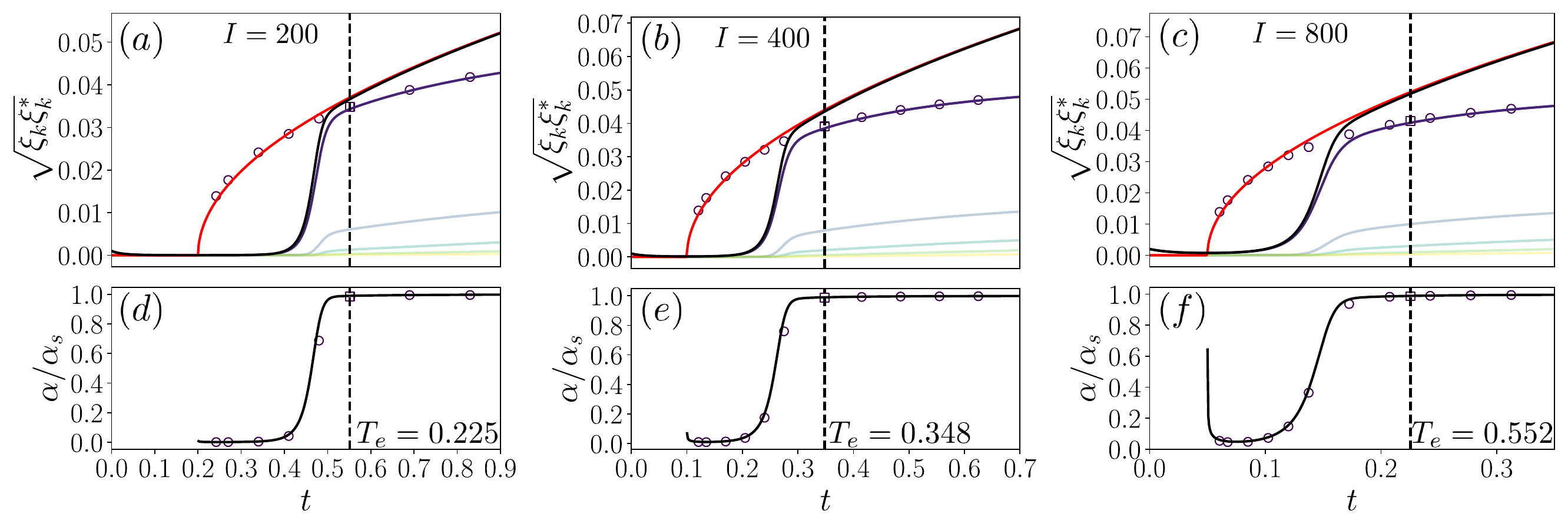}
    \caption{The delay time prediction (marked by the vertical dashed line) from the multiple-time scale analysis, compared to the fully nonlinear simulations. In (\textit{a},\textit{b},\textit{c}), the black (resp.\ purple) curves show the leading mode's amplitude evolution from the multiple-time-scale analysis (resp.\ fully nonlinear simulations). The red curves (resp.\ purple circles) show the stationary solution for the corresponding $a_1(t)$ from the multiple-time-scale analysis (resp.\ fully nonlinear simulations). The amplitude ratio of the time-dependent evolution and the corresponding stationary solution is shown in (\textit{d},\textit{e},\textit{f}), with the black curve (resp.\ purple circles) denoting the ratio from the multiple-time-scale analysis (resp.\ fully nonlinear simulations).} 
    \label{fig:delayed prediction}
\end{figure*}

When $X^2\gg \frac{2}{-QI}(\varkappa+IT_{e})^3 e^{-(2\varkappa_0T_{e}+IT_{e}^2)}$, the effect of the initial perturbation amplitude is no longer important, and the delay time can be explicitly predicted by 
\begin{equation}
    T_{e} \approx \frac{\rho}{\sqrt{2I(1-\rho^2)}}-\frac{\varkappa_0}{I},
    \label{eq:T_1^e simple}
\end{equation}
or $\varkappa = \varkappa_0 + IT_{e} = \sqrt{I\rho^2/2(1-\rho^2)}$. As shown in Fig.~\ref{fig:bifur time}, for fixed $\varkappa_0$ and $I$, the delay time $T_{e}$ first decreases as the initial perturbation increases, and then starts to saturate (around $10^{-3}$) to the value determined by $\varkappa_0$ and $I$ only.
Note that $\varkappa_0$ and $I$ are controllable parameters corresponding to the external forces, and thus in the physical system, as long as the droplet is perturbed by a perceivable amplitude (say, $>0.1\%$ of its initial radius), the delay time can be explicitly computed/controlled via Eq.~\eqref{eq:T_1^e simple}.

Figure~\ref{fig:delayed prediction} shows that the delay time $T_{e}$ evaluated from Eq.~\eqref{eq:T1e eval}, based on the physical parameters and initial perturbation can predict the bifurcation delay. Further, it is evident that this prediction compares favorably with the delayed time observed in the multiple-time-scale analysis and the fully nonlinear simulations. For example, $T_{e}$ can be taken as the minimum time needed for the time-dependent evolution to saturate to a predictable stationary state. When $T_1<T_{e}$, the dynamics is governed by exponential growth or decay. Subsequently, the amplitude experiences limited growth constrained by nonlinearity. Finally, when $T_1>T_{e}$, the dynamics saturate to a state governed by the balance of nonlinearity and dispersion, and the interface evolution is determined by the quasistatic variation of the bifurcation parameter, \textit{i.e.} the system responds to the (slow) external forcing instantaneously.

\begin{figure}
    \centering
    \includegraphics[keepaspectratio=true,width=\columnwidth]{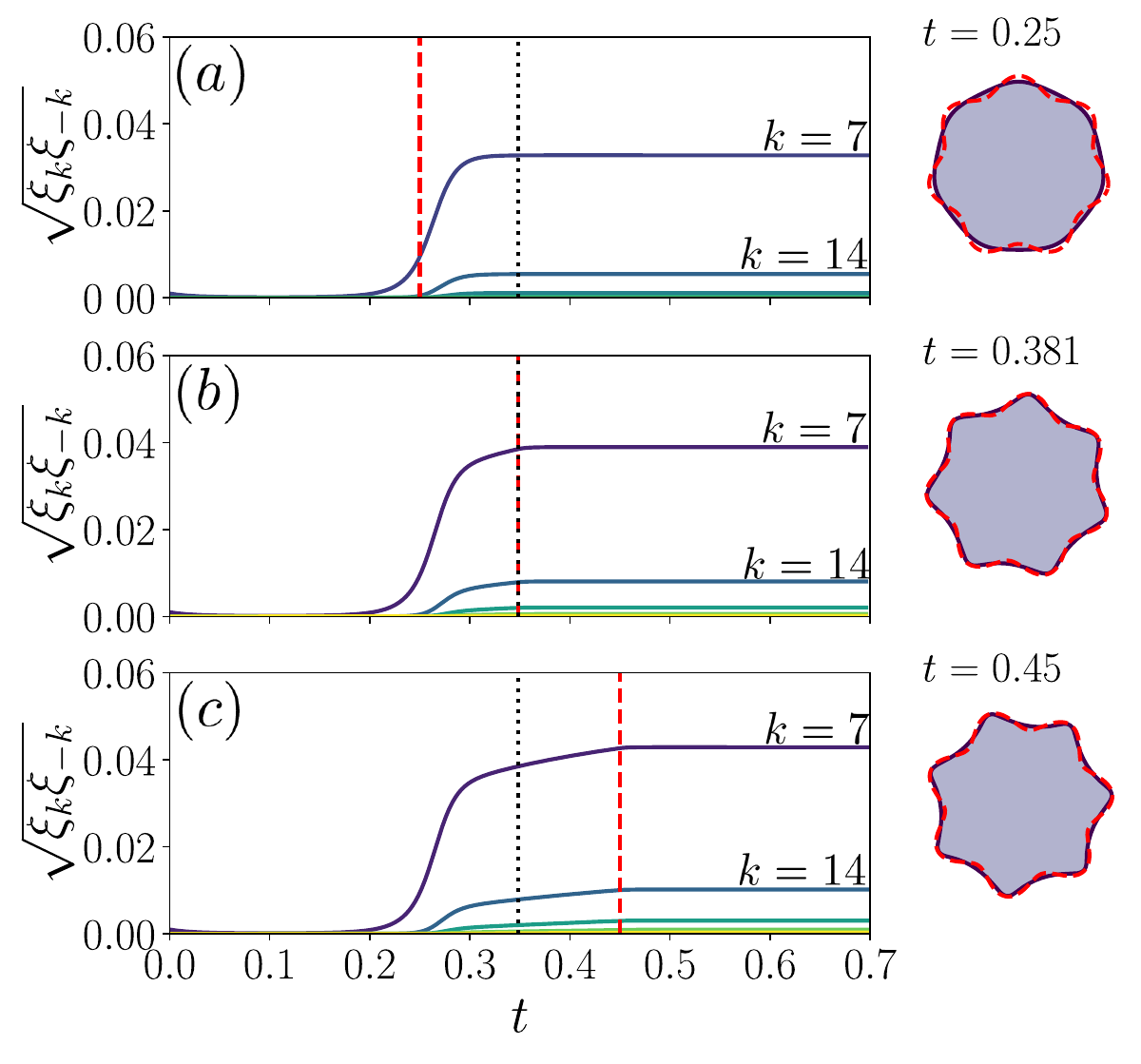}
    \caption{Control of the rotating droplet shapes (via the amplitude of the interfacial traveling wave) for (\textit{a}) $T_\mathrm{off}=0.25$, (\textit{b}) $T_\mathrm{off}=T_{e}\approx 0.348$, and (\textit{c}) $T_\mathrm{off}=0.45$. The curves show the leading mode's amplitude evolution from the fully nonlinear simulations. The red dashed vertical line denotes the turn-off time $T_\mathrm{off}$; the black dotted vertical line denotes the delay time prediction $T_{e}$. The colored droplets are the real-time profiles from the fully nonlinear simulations (up to the corresponding times), and the red dashed outlines show the targetted profiles evaluated via Eq.~\eqref{eq:alpha_s}. Here, $I=400$, $X=0.001$, $\varkappa_0=-40$, and $\nba=1$, which are also the values used to evaluate $T_e$. For the convenience of the comparison, $\epsilon=1$ is taken such that $T_1$ and $t$ can be plotted at the same time scale. All other parameters are determined through Eq.~\eqref{eq:Linear growth} and Appendix~\ref{sec:ai bi ci di}.} 
    \label{fig:stop control}
\end{figure}

Figure~\ref{fig:stop control} shows fully nonlinear simulation examples using the value of $T_{e}$ to control the droplet's evolution. The azimuthal field's strength is fixed via $\nba=1$, and the radial field's strength, set by $\nbr$, is determined through $a_1=\varkappa_0+It$ for $t\leq T_{\mathrm{off}}$. For $T_1>T_{\mathrm{off}}$, $\nbr(T_1)=\nbr(T_{\mathrm{off}})$, \textit{i.e.}, both fields are static. Figure~\ref{fig:stop control} shows three cases, for different values of $T_{\mathrm{off}}$ but with the same initial perturbation strength $X=0.001$, and the same physical parameters (corresponding to $I=400$, $\varkappa_0=-40$). In Fig.~\ref{fig:stop control}(\textit{a}) $T_{\mathrm{off}}<T_{e}$. In this case, the radial field's strength stops increasing when the droplet amplitude is still in the linear regime, so it grows exponentially. In Fig.~\ref{fig:stop control}(\textit{b},\textit{c}), $T_{\mathrm{off}}\geq T_{e}$, and the radial field's strength stops increasing when the droplet begins to settle into the permanent rotating state. Note that, while $T_{e}$ is calculated through the multiple-time-scale analysis (which is a reduced model involving only two harmonic modes), it can still effectively capture the saturated state from the fully nonlinear simulations. The delay time $T_{e}$ can be controlled via an external magnetic field, which allows targeting the shape of the droplet, by evaluating the quasistatic solution $\alpha_s=\sqrt{-\varkappa/Q}$ with the linear growth rate $\varkappa$ at the targetted time $T_{\mathrm{off}}$.

\begin{figure*}
    \centering
    \includegraphics[keepaspectratio=true,width=\textwidth]{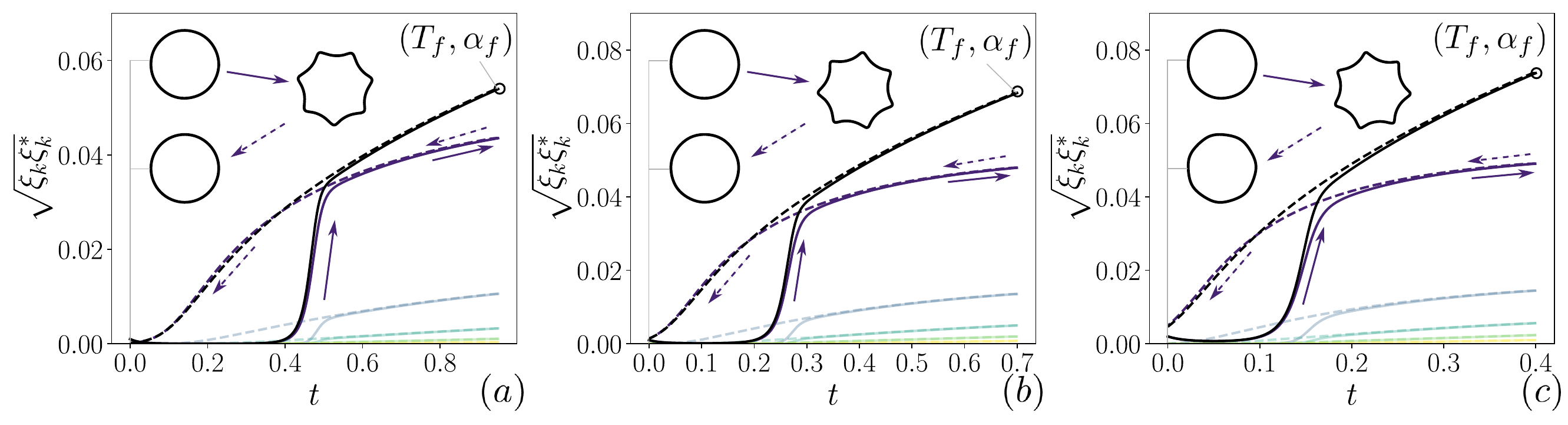}
    \caption{Dynamics under a reversed-time magnetic field: comparison of the prediction from the multiple-time-scale analysis and the fully nonlinear simulations. The solid (resp.\ dashed) curves show the forward (resp.\ reverse) process for which $\dot{\lambda}(k_f)>0$ ($\dot{\lambda}(k_f)<0$). The black (resp.\ purple) curve shows the leading mode amplitude evolution from the multiple-time scale analysis (resp.\ fully nonlinear simulations). The circle represents the ($T_f,\alpha_f$) state. In (\textit{a}), $T_f=0.95$, $X=0.001$, and $I=200$. In (\textit{b}), $T_f=0.7$, $X=0.001$, and $I=400$. In (\textit{c}), $T_f=0.4$, $X=0.002$, and $I=800$. In all three simulations, $\varkappa_0=-40$, and $\epsilon=1$ is taken such that $T_1$ and $t$ can be plotted at the same abscissa.} 
    \label{fig:Hysteresis}
\end{figure*}

\subsection{Irreversible dynamics under a time-reversed magnetic field} 

In the classic film \textit{Low Reynolds Number Flows} \cite{Taylor_NCFMF}, G.~I.~Taylor explained the physical meaning of reversibility --- ``low Reynolds number flows are reversible when the direction of motion of the boundaries, which gave rise to the flow, is reversed.'' The reversibility of Stokes flow is due to its steadiness and the fact that inertial forces are negligible. In this time-independent flow, the time-reversed problem solves the same equations as the original Stokes flow. These equations are linear in Taylor's example of Couette flow. The reversed fluid flow is the result of reversing the direction of the external forcing (rotation of the cylinder in the Couette flow example shown by Taylor). The reversibility is at first surprising, as it can be used to show that the initial state of the fluid is recovered under flow reversal, which in some ways may contradict intuition based on observations of everyday fluid flows.

In this study, the original problem is a Hele-Shaw flow, which in general is also expected to be reversible like a Stokes flow. Yet, the reversibility of the dynamics of the confined ferrofluid droplet is not an obvious consequence because nonlinearity arises from the surface forces (capillary tension and magnetic traction) acting on the fluid--fluid interface. The interface is also subjected to unsteady forcing by the time-dependent external magnetic field. And, thus, time-reversing the  magnetic field strengths does not return the fluid interface back to its initial shape. This irreversibility is demonstrated in Fig.~\ref{fig:Hysteresis}, in which the fully nonlinear simulations show the perturbation amplitude upon time-reversing the magnetic field can be (\textit{a}) smaller, (\textit{b}) similar, or (\textit{c}) larger than the initial perturbation.

The reversed process is initialized with the final state $(T_f,\alpha_f)$ from the forward process, then $\nba=1$ is fixed, and $\nbr$ is manipulated such that the linear growth rate decreases linearly. Specifically, $\Upsilon=\Re (a_1)=\Upsilon_0-IT_1$, where $\Upsilon_0=\varkappa_0+IT_f$. This protocol  achieves the reversal process of the external field, and $\Upsilon(t)=\varkappa(T_f-t)$, $\forall t \in [0,T_f]$. Note that, while the magnetic field is reversed, the external forces are not. The magnetic surface force depends on the interface's shape, and the irreversible evolution of the interface implies the irreversibility of the external forces in this problem. Thus, it is of interest to determine how to evaluate $\Upsilon(T_f)$, if the initial state corresponding to $\varkappa(0)$ cannot be fully recovered in this  irreversible system.

To answer this question, we first utilize  Eq.~\eqref{eq:Landau} from the multiple-time-scale analysis to formulate the reverse problem as:
\begin{equation}
    \frac{d\alpha}{dT_1} =(\Upsilon_0-IT_1)\alpha +Q\alpha^3.
    \label{eq:reverse}
\end{equation}
The solution for $\alpha$ can be calculated explicitly from
\begin{equation}
S(T_{1,a})\frac{1}{(\alpha_{a})^{2}}+2Q \tilde{S}(T_{1,a})=
S(T_{1,b})\frac{1}{(\alpha_{b})^{2}}+2Q \tilde{S}(T_{1,b}), 
\end{equation}
where the subscripts $a,b$ denote two arbitrary states and
\begin{subequations}\begin{align}
    S(t)&=e^{2\Upsilon_0 t -It^2},\\
    \tilde{S}(t) &= \int_0^t S(t')\, dt' = \frac{1}{2}\sqrt{\frac{\pi}{I}} e^{\Upsilon_0^2/I} 
    \erf\left(\frac{It-\Upsilon_0}{\sqrt{I}}\right).
\end{align}\end{subequations}
Taking $T_{1,a}=0$, $\alpha_a=\alpha_f$, then the reversed initial amplitude $\alpha_b$ can be predicted at $T_{1,a}=T_f$. One quick approximation can be made on finding the final state $(T_f,\alpha_f)$ when the forward evolution~\eqref{eq:forward landau} enters the quasistatic region, \textit{i.e.}, $T_f>T_{e}$, and the amplitude $\alpha_f$ can be approximated by Eq.~\eqref{eq:alpha_s}. 

Figure~\ref{fig:Hysteresis} shows a comparison of the forward and reversed processes approximated by Eqs.~\eqref{eq:forward landau} and~\eqref{eq:reverse}. In the quasistatic region, the evolution is close to reversible, after which the reverse evolution does not experience a sudden decrease in amplitude, which would parallel the rapid increase during the forward process. The cycling process under the fully nonlinear simulation shows similar dynamics. It is interesting to note that $\alpha_f$ from the simulation and Eq.~\eqref{eq:forward landau} are different at $t=T_f$, yet the predictions of both during the reverse process eventually coincide in the small-$t$ region, meaning that Eq.~\eqref{eq:reverse} provides a good approximation to the reversed ``initial'' amplitude.

{This result is very similar to the one reported in the experimental work~\cite{BS84}, wherein the peak of a magnetic fluid interface attains different amplitudes at the same field strength upon cycling the external magnetic field. This effect was attributed to the strong permeability of the ferrofluid. While in our work, the hysteresis-like behavior is mainly due to the time-dependent field's interaction with the interfacial nonlinearity, which is captured by the reduced models in Eq.~\eqref{eq:forward landau} and Eq.~\eqref{eq:reverse}. The difference between these evolution equations highlights the hysteresis-like behavior.}

On the one hand, Eq.~\eqref{eq:reverse} provides a tool for predicting the time-reversed process. On the other hand, this equation also provides a new point of view on the observed irreversibility. Solutions to Eq.~\eqref{eq:forward landau} in the $(T_1,\alpha)$ plane, and 
solutions to Eq.~\eqref{eq:reverse} in the $(T_f-T_1,\alpha)$ plane are two families of curves that intersect at $(T_f,\alpha_f)$. The initial condition~\eqref{eq:forward landau} determines a certain curve in the forward family, along which any arbitrary $(T_f,\alpha_f)$ can be found as the intersection point with the curve in the reverse family determined by Eq.~\eqref{eq:reverse}. Importantly, these two curves intersect only at $(T_f,\alpha_f)$ and do not overlap.

\section{Discussion and conclusion}
\label{sec:conclusion}
Previously, we demonstrated that the combination of static radial and azimuthal magnetic fields deforms a ferrofluid droplet confined in a Hele-Shaw cell into a stably spinning ``gear,'' whose rotation is driven by interfacial waves \cite{YC21}. In this study, we show that a periodic traveling wave on the droplet's interface is stable, and its dynamics is governed by a Hopf bifurcation at the critical growth rate. A center manifold reduction shows the geometrical equivalence between a two-harmonic-mode coupled ODE system describing the interface evolution and a supercritical Hopf bifurcation. This reduction is supported by the amplitude (Landau) equation derived from a multiple-time scale analysis, which also reveals how the marginally unstable linear solution is equilibrated by weak nonlinearity. Both methods adequately predict the fully nonlinear evolution, as demonstrated by comparisons between the theory and fully nonlinear, interface-resolved simulations of the original PDE system. 

{The intrinsic reason why a simple, local ODE can  approximate the fully nonlocal dynamics is discussed, also in the context of the static problem considered in \cite{AOC04}. However, unlike the case in \cite{AOC04}, we are unable to obtain a single curvature ODE for the dynamic problem, due to the difficulty of eliminating the nonlocal term from the vortex-sheet formulation of the full Hele-Shaw problem. This task remains an open question, specifically whether such a single curvature ODE even exists to exactly describe the family of traveling wave solutions discussed herein. To further understand that challenge, suppose that vortex elements on the interface are subjected to rigid rotation. In this case, a moving frame transformation would eliminate the relative velocity (and, thus, the nonlocal term). However, to perform a moving frame transformation, the exact traveling wave  velocity needs to be found, which is still nontrivial. On the other hand, if the interface is not rotating as a rigid body, then the elements on the interface have some local rotation rate, which collectively leads to the interfacial wave. In this case, when the local velocity is nonuniformly distributed along the interface, a moving frame transformation may not exist. The success of the approximations in the present work might imply the existence of such a curvature equation, but how to obtain it is left as an open question. Answering this question would surely provide further examples of the relevance of elastica solutions.} 

Next, with the reduced model revealing the key dynamical features, we designed a slowly-varying radial magnetic field such that the timing of the emergence of the spinning ``gear'' can be controlled. This work is inspired by the well-known delay behavior of dynamic Hopf bifurcations. In this study, the delay time is predicted based on the fact that the time-varying amplitude equation finally saturates to the quasistatic amplitude. This time can be manipulated purely via an external magnetic field by controlling the linear growth rate and its rate of change. We also studied the evolution under a time-reversed magnetic field. While we found that the evolution of the droplet is irreversible due to the nonlinearity in the interface condition, the reverse evolution, and the final stated achieved under it, can still be well approximated by the reversed amplitude equation. 

In this work, the bifurcation parameter is controlled by a simple linear variation, which allows for the explicit analytical solution of the amplitude equation, and the approximation of the delay time. 
{The linear variation with time is expected to be the simplest strategy that can be realized in experiments, as it only requires increasing the magnetic field strength at a constant rate. Thus,  by explicitly predicting the delay time, our work enables the effective design of the target control. Further, the selection of a linear variation scheme requires minimal algebraic calculations to obtain a straightforward prediction. 
}
Other control protocols, such as periodic forcing, can also be considered, providing a different view on the accumulation of the time-dependent evolution. 
{For example, a log-varying, an exponentially increasing, and an oscillating time-dependent protocol are highlighted in Appendix~\ref{sec:protocols}, which may form the basis of further explorations.}
The proposed reduction method can be generally applied to other interfacial problems governed by a finite number of harmonic modes. Our mode-reduction approach also allows for the effective and computationally inexpensive prediction of the dynamics, as well as for ``reverse engineering'' of time-dependent forcing schemes (\textit{i.e.}, choosing a forcing that generates dynamics of interest), such as those aiming to achieve pattern stabilization~\cite{zks2015} or self-similar evolution~\cite{LLFP09,azll22} of fluids confined in Hele-Shaw cells. 

\begin{acknowledgements} 
This research was supported by the U.S.\ National Science Foundation under grant no.~CMMI-2029540 (to I.C.C.) and a Bilsland Dissertation Fellowship from The Graduate School at Purdue University (to Z.Y.). I.C.C.\ would also like to acknowledge the hospitality of the University of Nicosia, Cyprus, where this work was completed thanks to a Fulbright U.S.\ Scholar award from the U.S.\ Department of State.
\end{acknowledgements}

\bibliography{Mendeley_reference,yu_reference}

\begin{thebibliography}{61}%
\makeatletter
\providecommand \@ifxundefined [1]{%
 \@ifx{#1\undefined}
}%
\providecommand \@ifnum [1]{%
 \ifnum #1\expandafter \@firstoftwo
 \else \expandafter \@secondoftwo
 \fi
}%
\providecommand \@ifx [1]{%
 \ifx #1\expandafter \@firstoftwo
 \else \expandafter \@secondoftwo
 \fi
}%
\providecommand \natexlab [1]{#1}%
\providecommand \enquote  [1]{``#1''}%
\providecommand \bibnamefont  [1]{#1}%
\providecommand \bibfnamefont [1]{#1}%
\providecommand \citenamefont [1]{#1}%
\providecommand \href@noop [0]{\@secondoftwo}%
\providecommand \href [0]{\begingroup \@sanitize@url \@href}%
\providecommand \@href[1]{\@@startlink{#1}\@@href}%
\providecommand \@@href[1]{\endgroup#1\@@endlink}%
\providecommand \@sanitize@url [0]{\catcode `\\12\catcode `\$12\catcode
  `\&12\catcode `\#12\catcode `\^12\catcode `\_12\catcode `\%12\relax}%
\providecommand \@@startlink[1]{}%
\providecommand \@@endlink[0]{}%
\providecommand \url  [0]{\begingroup\@sanitize@url \@url }%
\providecommand \@url [1]{\endgroup\@href {#1}{\urlprefix }}%
\providecommand \urlprefix  [0]{URL }%
\providecommand \Eprint [0]{\href }%
\providecommand \doibase [0]{https://doi.org/}%
\providecommand \selectlanguage [0]{\@gobble}%
\providecommand \bibinfo  [0]{\@secondoftwo}%
\providecommand \bibfield  [0]{\@secondoftwo}%
\providecommand \translation [1]{[#1]}%
\providecommand \BibitemOpen [0]{}%
\providecommand \bibitemStop [0]{}%
\providecommand \bibitemNoStop [0]{.\EOS\space}%
\providecommand \EOS [0]{\spacefactor3000\relax}%
\providecommand \BibitemShut  [1]{\csname bibitem#1\endcsname}%
\let\auto@bib@innerbib\@empty
\bibitem [{\citenamefont {Shliomis}(1974)}]{S74}%
  \BibitemOpen
  \bibfield  {author} {\bibinfo {author} {\bibfnamefont {M.~I.}\ \bibnamefont
  {Shliomis}},\ }\bibfield  {title} {\bibinfo {title} {{Magnetic fluids}},\
  }\href {https://doi.org/10.1070/PU1974v017n02ABEH004332} {\bibfield
  {journal} {\bibinfo  {journal} {Sov. Phys. Usp.}\ }\textbf {\bibinfo {volume}
  {17}},\ \bibinfo {pages} {153} (\bibinfo {year} {1974})}\BibitemShut
  {NoStop}%
\bibitem [{\citenamefont {Rosensweig}(1987)}]{R87}%
  \BibitemOpen
  \bibfield  {author} {\bibinfo {author} {\bibfnamefont {R.~E.}\ \bibnamefont
  {Rosensweig}},\ }\bibfield  {title} {\bibinfo {title} {{Magnetic Fluids}},\
  }\href {https://doi.org/10.1146/annurev.fl.19.010187.002253} {\bibfield
  {journal} {\bibinfo  {journal} {Annu. Rev. Fluid Mech.}\ }\textbf {\bibinfo
  {volume} {19}},\ \bibinfo {pages} {437} (\bibinfo {year} {1987})}\BibitemShut
  {NoStop}%
\bibitem [{\citenamefont {Rosensweig}(2014)}]{R13_Ferrohydrodynamics}%
  \BibitemOpen
  \bibfield  {author} {\bibinfo {author} {\bibfnamefont {R.}~\bibnamefont
  {Rosensweig}},\ }\href@noop {} {\emph {\bibinfo {title}
  {Ferrohydrodynamics}}}\ (\bibinfo  {publisher} {Dover Publications},\
  \bibinfo {address} {Mineola, NY},\ \bibinfo {year} {2014})\ \bibinfo {note}
  {republication of the 1997 edition}\BibitemShut {NoStop}%
\bibitem [{\citenamefont {Blums}\ \emph {et~al.}(1997)\citenamefont {Blums},
  \citenamefont {Cebers},\ and\ \citenamefont {Maiorov}}]{BCM10}%
  \BibitemOpen
  \bibfield  {author} {\bibinfo {author} {\bibfnamefont {E.}~\bibnamefont
  {Blums}}, \bibinfo {author} {\bibfnamefont {A.}~\bibnamefont {Cebers}},\ and\
  \bibinfo {author} {\bibfnamefont {M.}~\bibnamefont {Maiorov}},\ }\href
  {https://doi.org/10.1515/9783110807356} {\emph {\bibinfo {title} {Magnetic
  Fluids}}}\ (\bibinfo  {publisher} {De Gruyter},\ \bibinfo {address}
  {Berlin},\ \bibinfo {year} {1997})\BibitemShut {NoStop}%
\bibitem [{\citenamefont {Huang}\ and\ \citenamefont {Michels}(2020)}]{HM20}%
  \BibitemOpen
  \bibfield  {author} {\bibinfo {author} {\bibfnamefont {L.}~\bibnamefont
  {Huang}}\ and\ \bibinfo {author} {\bibfnamefont {D.~L.}\ \bibnamefont
  {Michels}},\ }\bibfield  {title} {\bibinfo {title} {Surface-only
  ferrofluids},\ }\href {https://doi.org/10.1145/3414685.3417799} {\bibfield
  {journal} {\bibinfo  {journal} {ACM Trans. Graph.}\ }\textbf {\bibinfo
  {volume} {39}},\ \bibinfo {pages} {174} (\bibinfo {year} {2020})}\BibitemShut
  {NoStop}%
\bibitem [{\citenamefont {Voltairas}\ \emph {et~al.}(2002)\citenamefont
  {Voltairas}, \citenamefont {Fotiadis},\ and\ \citenamefont
  {Michalis}}]{VFM02}%
  \BibitemOpen
  \bibfield  {author} {\bibinfo {author} {\bibfnamefont {P.}~\bibnamefont
  {Voltairas}}, \bibinfo {author} {\bibfnamefont {D.}~\bibnamefont
  {Fotiadis}},\ and\ \bibinfo {author} {\bibfnamefont {L.}~\bibnamefont
  {Michalis}},\ }\bibfield  {title} {\bibinfo {title} {Hydrodynamics of
  magnetic drug targeting},\ }\href
  {https://doi.org/10.1016/S0021-9290(02)00034-9} {\bibfield  {journal}
  {\bibinfo  {journal} {J. Biomech.}\ }\textbf {\bibinfo {volume} {35}},\
  \bibinfo {pages} {813} (\bibinfo {year} {2002})}\BibitemShut {NoStop}%
\bibitem [{\citenamefont {Serwane}\ \emph {et~al.}(2017)\citenamefont
  {Serwane}, \citenamefont {Mongera}, \citenamefont {Rowghanian}, \citenamefont
  {Kealhofer}, \citenamefont {Lucio}, \citenamefont {Hockenbery},\ and\
  \citenamefont {Campas}}]{SMRK17}%
  \BibitemOpen
  \bibfield  {author} {\bibinfo {author} {\bibfnamefont {F.}~\bibnamefont
  {Serwane}}, \bibinfo {author} {\bibfnamefont {A.}~\bibnamefont {Mongera}},
  \bibinfo {author} {\bibfnamefont {P.}~\bibnamefont {Rowghanian}}, \bibinfo
  {author} {\bibfnamefont {D.~A.}\ \bibnamefont {Kealhofer}}, \bibinfo {author}
  {\bibfnamefont {A.~A.}\ \bibnamefont {Lucio}}, \bibinfo {author}
  {\bibfnamefont {Z.~M.}\ \bibnamefont {Hockenbery}},\ and\ \bibinfo {author}
  {\bibfnamefont {O.}~\bibnamefont {Campas}},\ }\bibfield  {title} {\bibinfo
  {title} {{\it In vivo} quantification of spatially varying mechanical
  properties in developing tissues},\ }\href
  {https://doi.org/10.1038/nmeth.4101} {\bibfield  {journal} {\bibinfo
  {journal} {Nat. Methods}\ }\textbf {\bibinfo {volume} {14}},\ \bibinfo
  {pages} {181} (\bibinfo {year} {2017})}\BibitemShut {NoStop}%
\bibitem [{\citenamefont {Fan}\ \emph {et~al.}(2020)\citenamefont {Fan},
  \citenamefont {Dong}, \citenamefont {Karacakol}, \citenamefont {Xie},\ and\
  \citenamefont {Sitti}}]{FDKXS20}%
  \BibitemOpen
  \bibfield  {author} {\bibinfo {author} {\bibfnamefont {X.}~\bibnamefont
  {Fan}}, \bibinfo {author} {\bibfnamefont {X.}~\bibnamefont {Dong}}, \bibinfo
  {author} {\bibfnamefont {A.~C.}\ \bibnamefont {Karacakol}}, \bibinfo {author}
  {\bibfnamefont {H.}~\bibnamefont {Xie}},\ and\ \bibinfo {author}
  {\bibfnamefont {M.}~\bibnamefont {Sitti}},\ }\bibfield  {title} {\bibinfo
  {title} {Reconfigurable multifunctional ferrofluid droplet robots},\ }\href
  {https://doi.org/10.1073/pnas.2016388117} {\bibfield  {journal} {\bibinfo
  {journal} {Proc. Natl Acad. Sci. USA}\ }\textbf {\bibinfo {volume} {117}},\
  \bibinfo {pages} {27916} (\bibinfo {year} {2020})}\BibitemShut {NoStop}%
\bibitem [{\citenamefont {Ahmed}\ \emph {et~al.}(2021)\citenamefont {Ahmed},
  \citenamefont {Ilami}, \citenamefont {Bant}, \citenamefont {Beigzadeh},\ and\
  \citenamefont {Marvi}}]{AIBBM20}%
  \BibitemOpen
  \bibfield  {author} {\bibinfo {author} {\bibfnamefont {R.}~\bibnamefont
  {Ahmed}}, \bibinfo {author} {\bibfnamefont {M.}~\bibnamefont {Ilami}},
  \bibinfo {author} {\bibfnamefont {J.}~\bibnamefont {Bant}}, \bibinfo {author}
  {\bibfnamefont {B.}~\bibnamefont {Beigzadeh}},\ and\ \bibinfo {author}
  {\bibfnamefont {H.}~\bibnamefont {Marvi}},\ }\bibfield  {title} {\bibinfo
  {title} {A shapeshifting ferrofluidic robot},\ }\href
  {https://doi.org/10.1089/soro.2019.0184} {\bibfield  {journal} {\bibinfo
  {journal} {Soft Robotics}\ }\textbf {\bibinfo {volume} {8}},\ \bibinfo
  {pages} {687} (\bibinfo {year} {2021})}\BibitemShut {NoStop}%
\bibitem [{\citenamefont {Yu}\ \emph {et~al.}(2020)\citenamefont {Yu},
  \citenamefont {Lin}, \citenamefont {Wang}, \citenamefont {He}, \citenamefont
  {Chen}, \citenamefont {Sun}, \citenamefont {Lo}, \citenamefont {Cheng},
  \citenamefont {Yeung}, \citenamefont {Tan}, \citenamefont {Carlo},\ and\
  \citenamefont {Emaminejad}}]{YLW20}%
  \BibitemOpen
  \bibfield  {author} {\bibinfo {author} {\bibfnamefont {W.}~\bibnamefont
  {Yu}}, \bibinfo {author} {\bibfnamefont {H.}~\bibnamefont {Lin}}, \bibinfo
  {author} {\bibfnamefont {Y.}~\bibnamefont {Wang}}, \bibinfo {author}
  {\bibfnamefont {X.}~\bibnamefont {He}}, \bibinfo {author} {\bibfnamefont
  {N.}~\bibnamefont {Chen}}, \bibinfo {author} {\bibfnamefont {K.}~\bibnamefont
  {Sun}}, \bibinfo {author} {\bibfnamefont {D.}~\bibnamefont {Lo}}, \bibinfo
  {author} {\bibfnamefont {B.}~\bibnamefont {Cheng}}, \bibinfo {author}
  {\bibfnamefont {C.}~\bibnamefont {Yeung}}, \bibinfo {author} {\bibfnamefont
  {J.}~\bibnamefont {Tan}}, \bibinfo {author} {\bibfnamefont {D.~D.}\
  \bibnamefont {Carlo}},\ and\ \bibinfo {author} {\bibfnamefont
  {S.}~\bibnamefont {Emaminejad}},\ }\bibfield  {title} {\bibinfo {title} {A
  ferrobotic system for automated microfluidic logistics},\ }\href
  {https://doi.org/10.1126/scirobotics.aba4411} {\bibfield  {journal} {\bibinfo
   {journal} {Sci. Robot.}\ }\textbf {\bibinfo {volume} {5}},\ \bibinfo {pages}
  {eaba4411} (\bibinfo {year} {2020})}\BibitemShut {NoStop}%
\bibitem [{\citenamefont {Hele-Shaw}(1898)}]{HS98}%
  \BibitemOpen
  \bibfield  {author} {\bibinfo {author} {\bibfnamefont {H.~S.}\ \bibnamefont
  {Hele-Shaw}},\ }\bibfield  {title} {\bibinfo {title} {{The flow of water}},\
  }\href {https://doi.org/10.1038/058034a0} {\bibfield  {journal} {\bibinfo
  {journal} {Nature}\ }\textbf {\bibinfo {volume} {58}},\ \bibinfo {pages} {34}
  (\bibinfo {year} {1898})}\BibitemShut {NoStop}%
\bibitem [{\citenamefont {Bensimon}\ \emph {et~al.}(1986)\citenamefont
  {Bensimon}, \citenamefont {Kadanoff}, \citenamefont {Liang}, \citenamefont
  {Shraiman},\ and\ \citenamefont {Tang}}]{BKLST86}%
  \BibitemOpen
  \bibfield  {author} {\bibinfo {author} {\bibfnamefont {D.}~\bibnamefont
  {Bensimon}}, \bibinfo {author} {\bibfnamefont {L.~P.}\ \bibnamefont
  {Kadanoff}}, \bibinfo {author} {\bibfnamefont {S.}~\bibnamefont {Liang}},
  \bibinfo {author} {\bibfnamefont {B.~I.}\ \bibnamefont {Shraiman}},\ and\
  \bibinfo {author} {\bibfnamefont {C.}~\bibnamefont {Tang}},\ }\bibfield
  {title} {\bibinfo {title} {{Viscous flows in two dimensions}},\ }\href
  {https://doi.org/10.1103/RevModPhys.58.977} {\bibfield  {journal} {\bibinfo
  {journal} {Rev. Mod. Phys.}\ }\textbf {\bibinfo {volume} {58}},\ \bibinfo
  {pages} {977} (\bibinfo {year} {1986})}\BibitemShut {NoStop}%
\bibitem [{\citenamefont {Rosensweig}\ \emph {et~al.}(1983)\citenamefont
  {Rosensweig}, \citenamefont {Zahn},\ and\ \citenamefont {Shumovich}}]{RZS83}%
  \BibitemOpen
  \bibfield  {author} {\bibinfo {author} {\bibfnamefont {R.~E.}\ \bibnamefont
  {Rosensweig}}, \bibinfo {author} {\bibfnamefont {M.}~\bibnamefont {Zahn}},\
  and\ \bibinfo {author} {\bibfnamefont {R.}~\bibnamefont {Shumovich}},\
  }\bibfield  {title} {\bibinfo {title} {{Labyrinthine instability in magnetic
  and dielectric fluids}},\ }\href
  {https://doi.org/10.1016/0304-8853(83)90416-X} {\bibfield  {journal}
  {\bibinfo  {journal} {J. Magn. Magn. Mater.}\ }\textbf {\bibinfo {volume}
  {39}},\ \bibinfo {pages} {127} (\bibinfo {year} {1983})}\BibitemShut
  {NoStop}%
\bibitem [{\citenamefont {Langer}\ \emph {et~al.}(1992)\citenamefont {Langer},
  \citenamefont {Goldstein},\ and\ \citenamefont {Jackson}}]{LGJ92}%
  \BibitemOpen
  \bibfield  {author} {\bibinfo {author} {\bibfnamefont {S.~A.}\ \bibnamefont
  {Langer}}, \bibinfo {author} {\bibfnamefont {R.~E.}\ \bibnamefont
  {Goldstein}},\ and\ \bibinfo {author} {\bibfnamefont {D.~P.}\ \bibnamefont
  {Jackson}},\ }\bibfield  {title} {\bibinfo {title} {Dynamics of labyrinthine
  pattern formation in magnetic fluids},\ }\href
  {https://doi.org/10.1103/PhysRevA.46.4894} {\bibfield  {journal} {\bibinfo
  {journal} {Phys. Rev. A}\ }\textbf {\bibinfo {volume} {46}},\ \bibinfo
  {pages} {4894} (\bibinfo {year} {1992})}\BibitemShut {NoStop}%
\bibitem [{\citenamefont {Oliveira}\ \emph {et~al.}(2008)\citenamefont
  {Oliveira}, \citenamefont {Miranda},\ and\ \citenamefont {Leandro}}]{OML08}%
  \BibitemOpen
  \bibfield  {author} {\bibinfo {author} {\bibfnamefont {R.~M.}\ \bibnamefont
  {Oliveira}}, \bibinfo {author} {\bibfnamefont {J.~A.}\ \bibnamefont
  {Miranda}},\ and\ \bibinfo {author} {\bibfnamefont {E.~S.~G.}\ \bibnamefont
  {Leandro}},\ }\bibfield  {title} {\bibinfo {title} {Ferrofluid patterns in a
  radial magnetic field: Linear stability, nonlinear dynamics, and exact
  solutions},\ }\href {https://doi.org/10.1103/PhysRevE.77.016304} {\bibfield
  {journal} {\bibinfo  {journal} {Phys. Rev. E}\ }\textbf {\bibinfo {volume}
  {77}},\ \bibinfo {pages} {016304} (\bibinfo {year} {2008})}\BibitemShut
  {NoStop}%
\bibitem [{\citenamefont {Lira}\ and\ \citenamefont {Miranda}(2016)}]{LM16}%
  \BibitemOpen
  \bibfield  {author} {\bibinfo {author} {\bibfnamefont {S.~A.}\ \bibnamefont
  {Lira}}\ and\ \bibinfo {author} {\bibfnamefont {J.~A.}\ \bibnamefont
  {Miranda}},\ }\bibfield  {title} {\bibinfo {title} {{Ferrofluid patterns in
  Hele-Shaw cells: Exact, stable, stationary shape solutions}},\ }\href
  {https://doi.org/10.1103/PhysRevE.93.013129} {\bibfield  {journal} {\bibinfo
  {journal} {Phys. Rev. E}\ }\textbf {\bibinfo {volume} {93}},\ \bibinfo
  {pages} {013129} (\bibinfo {year} {2016})}\BibitemShut {NoStop}%
\bibitem [{\citenamefont {Lira}\ \emph {et~al.}(2010)\citenamefont {Lira},
  \citenamefont {Miranda},\ and\ \citenamefont {Oliveira}}]{LMO10}%
  \BibitemOpen
  \bibfield  {author} {\bibinfo {author} {\bibfnamefont {S.~A.}\ \bibnamefont
  {Lira}}, \bibinfo {author} {\bibfnamefont {J.~A.}\ \bibnamefont {Miranda}},\
  and\ \bibinfo {author} {\bibfnamefont {R.~M.}\ \bibnamefont {Oliveira}},\
  }\bibfield  {title} {\bibinfo {title} {Stationary shapes of confined rotating
  magnetic liquid droplets},\ }\href
  {https://doi.org/10.1103/PhysRevE.82.036318} {\bibfield  {journal} {\bibinfo
  {journal} {Phys. Rev. E}\ }\textbf {\bibinfo {volume} {82}},\ \bibinfo
  {pages} {036318} (\bibinfo {year} {2010})}\BibitemShut {NoStop}%
\bibitem [{\citenamefont {Dias}\ and\ \citenamefont {Miranda}(2015)}]{DM15}%
  \BibitemOpen
  \bibfield  {author} {\bibinfo {author} {\bibfnamefont {E.~O.}\ \bibnamefont
  {Dias}}\ and\ \bibinfo {author} {\bibfnamefont {J.~A.}\ \bibnamefont
  {Miranda}},\ }\bibfield  {title} {\bibinfo {title} {Azimuthal field
  instability in a confined ferrofluid},\ }\href
  {https://doi.org/10.1103/PhysRevE.91.023020} {\bibfield  {journal} {\bibinfo
  {journal} {Phys. Rev. E}\ }\textbf {\bibinfo {volume} {91}},\ \bibinfo
  {pages} {023020} (\bibinfo {year} {2015})}\BibitemShut {NoStop}%
\bibitem [{\citenamefont {Jackson}\ and\ \citenamefont {Miranda}(2007)}]{JM07}%
  \BibitemOpen
  \bibfield  {author} {\bibinfo {author} {\bibfnamefont {D.~P.}\ \bibnamefont
  {Jackson}}\ and\ \bibinfo {author} {\bibfnamefont {J.~A.}\ \bibnamefont
  {Miranda}},\ }\bibfield  {title} {\bibinfo {title} {{Confined ferrofluid
  droplet in crossed magnetic fields}},\ }\href
  {https://doi.org/10.1140/epje/i2007-10199-x} {\bibfield  {journal} {\bibinfo
  {journal} {Eur. Phys. J. E}\ }\textbf {\bibinfo {volume} {23}},\ \bibinfo
  {pages} {389} (\bibinfo {year} {2007})}\BibitemShut {NoStop}%
\bibitem [{\citenamefont {Yu}\ and\ \citenamefont
  {Christov}(2021{\natexlab{a}})}]{YC21}%
  \BibitemOpen
  \bibfield  {author} {\bibinfo {author} {\bibfnamefont {Z.}~\bibnamefont
  {Yu}}\ and\ \bibinfo {author} {\bibfnamefont {I.~C.}\ \bibnamefont
  {Christov}},\ }\bibfield  {title} {\bibinfo {title} {{Tuning a magnetic field
  to generate spinning ferrofluid droplets with controllable speed via
  nonlinear periodic interfacial waves}},\ }\href
  {https://doi.org/10.1103/PhysRevE.103.013103} {\bibfield  {journal} {\bibinfo
   {journal} {Phys. Rev. E}\ }\textbf {\bibinfo {volume} {103}},\ \bibinfo
  {pages} {013103} (\bibinfo {year} {2021}{\natexlab{a}})}\BibitemShut
  {NoStop}%
\bibitem [{\citenamefont {Oliveira}\ \emph {et~al.}(2021)\citenamefont
  {Oliveira}, \citenamefont {Coutinho}, \citenamefont {Anjos},\ and\
  \citenamefont {Miranda}}]{OCAM21}%
  \BibitemOpen
  \bibfield  {author} {\bibinfo {author} {\bibfnamefont {R.~M.}\ \bibnamefont
  {Oliveira}}, \bibinfo {author} {\bibfnamefont {I.~M.}\ \bibnamefont
  {Coutinho}}, \bibinfo {author} {\bibfnamefont {P.~H.~A.}\ \bibnamefont
  {Anjos}},\ and\ \bibinfo {author} {\bibfnamefont {J.~A.}\ \bibnamefont
  {Miranda}},\ }\bibfield  {title} {\bibinfo {title} {Shape instabilities in
  confined ferrofluids under crossed magnetic fields},\ }\href
  {https://doi.org/10.1103/PhysRevE.104.065113} {\bibfield  {journal} {\bibinfo
   {journal} {Phys. Rev. E}\ }\textbf {\bibinfo {volume} {104}},\ \bibinfo
  {pages} {065113} (\bibinfo {year} {2021})}\BibitemShut {NoStop}%
\bibitem [{\citenamefont {Livera}\ \emph {et~al.}(2022)\citenamefont {Livera},
  \citenamefont {Anjos},\ and\ \citenamefont {Miranda}}]{LAM22}%
  \BibitemOpen
  \bibfield  {author} {\bibinfo {author} {\bibfnamefont {P.~O.~S.}\
  \bibnamefont {Livera}}, \bibinfo {author} {\bibfnamefont {P.~H.~A.}\
  \bibnamefont {Anjos}},\ and\ \bibinfo {author} {\bibfnamefont {J.~A.}\
  \bibnamefont {Miranda}},\ }\bibfield  {title} {\bibinfo {title} {Ferrofluid
  annulus in crossed magnetic fields},\ }\href
  {https://doi.org/10.1103/PhysRevE.105.045106} {\bibfield  {journal} {\bibinfo
   {journal} {Phys. Rev. E}\ }\textbf {\bibinfo {volume} {105}},\ \bibinfo
  {pages} {045106} (\bibinfo {year} {2022})}\BibitemShut {NoStop}%
\bibitem [{\citenamefont {Coutinho}\ and\ \citenamefont
  {Miranda}(2022)}]{CM22}%
  \BibitemOpen
  \bibfield  {author} {\bibinfo {author} {\bibfnamefont {I.~M.}\ \bibnamefont
  {Coutinho}}\ and\ \bibinfo {author} {\bibfnamefont {J.~A.}\ \bibnamefont
  {Miranda}},\ }\bibfield  {title} {\bibinfo {title} {Field-controlled flow and
  shape of a magnetorheological fluid annulus},\ }\href
  {https://doi.org/10.1103/PhysRevE.106.025105} {\bibfield  {journal} {\bibinfo
   {journal} {Phys. Rev. E}\ }\textbf {\bibinfo {volume} {106}},\ \bibinfo
  {pages} {025105} (\bibinfo {year} {2022})}\BibitemShut {NoStop}%
\bibitem [{\citenamefont {Yu}\ and\ \citenamefont
  {Christov}(2021{\natexlab{b}})}]{YC21_rspa}%
  \BibitemOpen
  \bibfield  {author} {\bibinfo {author} {\bibfnamefont {Z.}~\bibnamefont
  {Yu}}\ and\ \bibinfo {author} {\bibfnamefont {I.~C.}\ \bibnamefont
  {Christov}},\ }\bibfield  {title} {\bibinfo {title} {Long-wave equation for a
  confined ferrofluid interface: periodic interfacial waves as dissipative
  solitons},\ }\href {https://doi.org/10.1098/rspa.2021.0550} {\bibfield
  {journal} {\bibinfo  {journal} {Proc. R. Soc. A}\ }\textbf {\bibinfo {volume}
  {477}},\ \bibinfo {pages} {20210550} (\bibinfo {year}
  {2021}{\natexlab{b}})}\BibitemShut {NoStop}%
\bibitem [{\citenamefont {Hopf}(1948)}]{hopf1948}%
  \BibitemOpen
  \bibfield  {author} {\bibinfo {author} {\bibfnamefont {E.}~\bibnamefont
  {Hopf}},\ }\bibfield  {title} {\bibinfo {title} {A mathematical example
  displaying features of turbulence},\ }\href
  {https://doi.org/10.1002/cpa.3160010401} {\bibfield  {journal} {\bibinfo
  {journal} {Commun. Pure Appl. Math.}\ }\textbf {\bibinfo {volume} {1}},\
  \bibinfo {pages} {303} (\bibinfo {year} {1948})}\BibitemShut {NoStop}%
\bibitem [{\citenamefont {Lorenz}(1963)}]{lorenz1963}%
  \BibitemOpen
  \bibfield  {author} {\bibinfo {author} {\bibfnamefont {E.~N.}\ \bibnamefont
  {Lorenz}},\ }\bibfield  {title} {\bibinfo {title} {Deterministic nonperiodic
  flow},\ }\href {https://doi.org/10.1175/1520-0469(1963)020<0130:DNF>2.0.CO;2}
  {\bibfield  {journal} {\bibinfo  {journal} {J. Atmos. Sci.}\ }\textbf
  {\bibinfo {volume} {20}},\ \bibinfo {pages} {130} (\bibinfo {year}
  {1963})}\BibitemShut {NoStop}%
\bibitem [{\citenamefont {Franco-G{\'o}mez}\ \emph {et~al.}(2018)\citenamefont
  {Franco-G{\'o}mez}, \citenamefont {Thompson}, \citenamefont {Hazel},\ and\
  \citenamefont {Juel}}]{FTHJ18}%
  \BibitemOpen
  \bibfield  {author} {\bibinfo {author} {\bibfnamefont {A.}~\bibnamefont
  {Franco-G{\'o}mez}}, \bibinfo {author} {\bibfnamefont {A.~B.}\ \bibnamefont
  {Thompson}}, \bibinfo {author} {\bibfnamefont {A.~L.}\ \bibnamefont
  {Hazel}},\ and\ \bibinfo {author} {\bibfnamefont {A.}~\bibnamefont {Juel}},\
  }\bibfield  {title} {\bibinfo {title} {Bubble propagation in {Hele-Shaw}
  channels with centred constrictions},\ }\href
  {https://doi.org/10.1088/1873-7005/aaa5cf} {\bibfield  {journal} {\bibinfo
  {journal} {Fluid Dyn. Res.}\ }\textbf {\bibinfo {volume} {50}},\ \bibinfo
  {pages} {021403} (\bibinfo {year} {2018})}\BibitemShut {NoStop}%
\bibitem [{\citenamefont {Keeler}\ \emph {et~al.}(2019)\citenamefont {Keeler},
  \citenamefont {Thompson}, \citenamefont {Lemoult}, \citenamefont {Juel},\
  and\ \citenamefont {Hazel}}]{KTLJH19}%
  \BibitemOpen
  \bibfield  {author} {\bibinfo {author} {\bibfnamefont {J.~S.}\ \bibnamefont
  {Keeler}}, \bibinfo {author} {\bibfnamefont {A.~B.}\ \bibnamefont
  {Thompson}}, \bibinfo {author} {\bibfnamefont {G.}~\bibnamefont {Lemoult}},
  \bibinfo {author} {\bibfnamefont {A.}~\bibnamefont {Juel}},\ and\ \bibinfo
  {author} {\bibfnamefont {A.~L.}\ \bibnamefont {Hazel}},\ }\bibfield  {title}
  {\bibinfo {title} {The influence of invariant solutions on the transient
  behaviour of an air bubble in a {Hele-Shaw} channel},\ }\href
  {https://doi.org/10.1098/rspa.2019.0434} {\bibfield  {journal} {\bibinfo
  {journal} {Proc. R. Soc. A}\ }\textbf {\bibinfo {volume} {475}},\ \bibinfo
  {pages} {20190434} (\bibinfo {year} {2019})}\BibitemShut {NoStop}%
\bibitem [{\citenamefont {Miranda}\ and\ \citenamefont {Widom}(1998)}]{MW98}%
  \BibitemOpen
  \bibfield  {author} {\bibinfo {author} {\bibfnamefont {J.~A.}\ \bibnamefont
  {Miranda}}\ and\ \bibinfo {author} {\bibfnamefont {M.}~\bibnamefont
  {Widom}},\ }\bibfield  {title} {\bibinfo {title} {Radial fingering in a
  {Hele-Shaw} cell: a weakly nonlinear analysis},\ }\href
  {https://doi.org/10.1016/S0167-2789(98)00097-9} {\bibfield  {journal}
  {\bibinfo  {journal} {Physica D}\ }\textbf {\bibinfo {volume} {120}},\
  \bibinfo {pages} {315} (\bibinfo {year} {1998})}\BibitemShut {NoStop}%
\bibitem [{\citenamefont {Miranda}\ and\ \citenamefont
  {Oliveira}(2004)}]{MO04}%
  \BibitemOpen
  \bibfield  {author} {\bibinfo {author} {\bibfnamefont {J.~A.}\ \bibnamefont
  {Miranda}}\ and\ \bibinfo {author} {\bibfnamefont {R.~M.}\ \bibnamefont
  {Oliveira}},\ }\bibfield  {title} {\bibinfo {title} {{Time-dependent gap
  Hele-Shaw cell with a ferrofluid: Evidence for an interfacial singularity
  inhibition by a magnetic field}},\ }\href
  {https://doi.org/10.1103/PhysRevE.69.066312} {\bibfield  {journal} {\bibinfo
  {journal} {Phys. Rev. E}\ }\textbf {\bibinfo {volume} {69}},\ \bibinfo
  {pages} {066312} (\bibinfo {year} {2004})}\BibitemShut {NoStop}%
\bibitem [{\citenamefont {Anjos}\ \emph {et~al.}(2018)\citenamefont {Anjos},
  \citenamefont {Lira},\ and\ \citenamefont {Miranda}}]{ALM18}%
  \BibitemOpen
  \bibfield  {author} {\bibinfo {author} {\bibfnamefont {P.~H.~A.}\
  \bibnamefont {Anjos}}, \bibinfo {author} {\bibfnamefont {S.~A.}\ \bibnamefont
  {Lira}},\ and\ \bibinfo {author} {\bibfnamefont {J.~A.}\ \bibnamefont
  {Miranda}},\ }\bibfield  {title} {\bibinfo {title} {{Fingering patterns in
  magnetic fluids: Perturbative solutions and the stability of exact stationary
  shapes}},\ }\href {https://doi.org/10.1103/PhysRevFluids.3.044002} {\bibfield
   {journal} {\bibinfo  {journal} {Phys. Rev. Fluids}\ }\textbf {\bibinfo
  {volume} {3}},\ \bibinfo {pages} {044002} (\bibinfo {year}
  {2018})}\BibitemShut {NoStop}%
\bibitem [{\citenamefont {Al-Housseiny}\ \emph {et~al.}(2012)\citenamefont
  {Al-Housseiny}, \citenamefont {Tsai},\ and\ \citenamefont {Stone}}]{ATS12}%
  \BibitemOpen
  \bibfield  {author} {\bibinfo {author} {\bibfnamefont {T.~T.}\ \bibnamefont
  {Al-Housseiny}}, \bibinfo {author} {\bibfnamefont {P.~A.}\ \bibnamefont
  {Tsai}},\ and\ \bibinfo {author} {\bibfnamefont {H.~A.}\ \bibnamefont
  {Stone}},\ }\bibfield  {title} {\bibinfo {title} {Control of interfacial
  instabilities using flow geometry},\ }\href
  {https://doi.org/10.1038/nphys2396} {\bibfield  {journal} {\bibinfo
  {journal} {Nat. Phys.}\ }\textbf {\bibinfo {volume} {8}},\ \bibinfo {pages}
  {747} (\bibinfo {year} {2012})}\BibitemShut {NoStop}%
\bibitem [{\citenamefont {Pihler-Puzovi{\'{c}}}\ \emph
  {et~al.}(2012)\citenamefont {Pihler-Puzovi{\'{c}}}, \citenamefont {Illien},
  \citenamefont {Heil},\ and\ \citenamefont {Juel}}]{PPIHJ12}%
  \BibitemOpen
  \bibfield  {author} {\bibinfo {author} {\bibfnamefont {D.}~\bibnamefont
  {Pihler-Puzovi{\'{c}}}}, \bibinfo {author} {\bibfnamefont {P.}~\bibnamefont
  {Illien}}, \bibinfo {author} {\bibfnamefont {M.}~\bibnamefont {Heil}},\ and\
  \bibinfo {author} {\bibfnamefont {A.}~\bibnamefont {Juel}},\ }\bibfield
  {title} {\bibinfo {title} {{Suppression of complex fingerlike patterns at the
  interface between air and a viscous fluid by elastic membranes}},\ }\href
  {https://doi.org/10.1103/PhysRevLett.108.074502} {\bibfield  {journal}
  {\bibinfo  {journal} {Phys. Rev. Lett.}\ }\textbf {\bibinfo {volume} {108}},\
  \bibinfo {pages} {074502} (\bibinfo {year} {2012})}\BibitemShut {NoStop}%
\bibitem [{\citenamefont {Mirzadeh}\ and\ \citenamefont {Bazant}(2017)}]{MB17}%
  \BibitemOpen
  \bibfield  {author} {\bibinfo {author} {\bibfnamefont {M.}~\bibnamefont
  {Mirzadeh}}\ and\ \bibinfo {author} {\bibfnamefont {M.~Z.}\ \bibnamefont
  {Bazant}},\ }\bibfield  {title} {\bibinfo {title} {{Electrokinetic Control of
  Viscous Fingering}},\ }\href {https://doi.org/10.1103/PhysRevLett.119.174501}
  {\bibfield  {journal} {\bibinfo  {journal} {Phys. Rev. Lett.}\ }\textbf
  {\bibinfo {volume} {119}},\ \bibinfo {pages} {174501} (\bibinfo {year}
  {2017})}\BibitemShut {NoStop}%
\bibitem [{\citenamefont {Miranda}(2000)}]{M00}%
  \BibitemOpen
  \bibfield  {author} {\bibinfo {author} {\bibfnamefont {J.~A.}\ \bibnamefont
  {Miranda}},\ }\bibfield  {title} {\bibinfo {title} {Rotating {Hele-Shaw}
  cells with ferrofluids},\ }\href {https://doi.org/10.1103/PhysRevE.62.2985}
  {\bibfield  {journal} {\bibinfo  {journal} {Phys. Rev. E}\ }\textbf {\bibinfo
  {volume} {62}},\ \bibinfo {pages} {2985} (\bibinfo {year}
  {2000})}\BibitemShut {NoStop}%
\bibitem [{\citenamefont {Morrow}\ \emph {et~al.}(2019)\citenamefont {Morrow},
  \citenamefont {Moroney},\ and\ \citenamefont {McCue}}]{MMM19}%
  \BibitemOpen
  \bibfield  {author} {\bibinfo {author} {\bibfnamefont {L.~C.}\ \bibnamefont
  {Morrow}}, \bibinfo {author} {\bibfnamefont {T.~J.}\ \bibnamefont
  {Moroney}},\ and\ \bibinfo {author} {\bibfnamefont {S.~W.}\ \bibnamefont
  {McCue}},\ }\bibfield  {title} {\bibinfo {title} {{Numerical investigation of
  controlling interfacial instabilities in non-standard Hele-Shaw
  configurations}},\ }\href {https://doi.org/10.1017/jfm.2019.623} {\bibfield
  {journal} {\bibinfo  {journal} {J. Fluid Mech.}\ }\textbf {\bibinfo {volume}
  {877}},\ \bibinfo {pages} {1063} (\bibinfo {year} {2019})}\BibitemShut
  {NoStop}%
\bibitem [{\citenamefont {Cardoso}\ and\ \citenamefont {Woods}(1995)}]{CW95}%
  \BibitemOpen
  \bibfield  {author} {\bibinfo {author} {\bibfnamefont {S.~S.~S.}\
  \bibnamefont {Cardoso}}\ and\ \bibinfo {author} {\bibfnamefont {A.~W.}\
  \bibnamefont {Woods}},\ }\bibfield  {title} {\bibinfo {title} {The formation
  of drops through viscous instability},\ }\href
  {https://doi.org/10.1017/S0022112095001364} {\bibfield  {journal} {\bibinfo
  {journal} {J. Fluid Mech.}\ }\textbf {\bibinfo {volume} {289}},\ \bibinfo
  {pages} {351} (\bibinfo {year} {1995})}\BibitemShut {NoStop}%
\bibitem [{\citenamefont {Li}\ \emph {et~al.}(2009)\citenamefont {Li},
  \citenamefont {Lowengrub}, \citenamefont {Fontana},\ and\ \citenamefont
  {Palffy-Muhoray}}]{LLFP09}%
  \BibitemOpen
  \bibfield  {author} {\bibinfo {author} {\bibfnamefont {S.}~\bibnamefont
  {Li}}, \bibinfo {author} {\bibfnamefont {J.~S.}\ \bibnamefont {Lowengrub}},
  \bibinfo {author} {\bibfnamefont {J.}~\bibnamefont {Fontana}},\ and\ \bibinfo
  {author} {\bibfnamefont {P.}~\bibnamefont {Palffy-Muhoray}},\ }\bibfield
  {title} {\bibinfo {title} {{Control of Viscous Fingering Patterns in a Radial
  Hele-Shaw Cell}},\ }\href {https://doi.org/10.1103/PhysRevLett.102.174501}
  {\bibfield  {journal} {\bibinfo  {journal} {Phys. Rev. Lett.}\ }\textbf
  {\bibinfo {volume} {102}},\ \bibinfo {pages} {174501} (\bibinfo {year}
  {2009})}\BibitemShut {NoStop}%
\bibitem [{\citenamefont {Zheng}\ \emph {et~al.}(2015)\citenamefont {Zheng},
  \citenamefont {Kim},\ and\ \citenamefont {Stone}}]{zks2015}%
  \BibitemOpen
  \bibfield  {author} {\bibinfo {author} {\bibfnamefont {Z.}~\bibnamefont
  {Zheng}}, \bibinfo {author} {\bibfnamefont {H.}~\bibnamefont {Kim}},\ and\
  \bibinfo {author} {\bibfnamefont {H.~A.}\ \bibnamefont {Stone}},\ }\bibfield
  {title} {\bibinfo {title} {{Controlling viscous fingering using
  time-dependent strategies}},\ }\href
  {https://doi.org/10.1103/PhysRevLett.115.174501} {\bibfield  {journal}
  {\bibinfo  {journal} {Phys. Rev. Lett.}\ }\textbf {\bibinfo {volume} {115}},\
  \bibinfo {pages} {174501} (\bibinfo {year} {2015})}\BibitemShut {NoStop}%
\bibitem [{\citenamefont {Shelley}\ \emph {et~al.}(1997)\citenamefont
  {Shelley}, \citenamefont {Tian},\ and\ \citenamefont {Wlodarski}}]{STW97}%
  \BibitemOpen
  \bibfield  {author} {\bibinfo {author} {\bibfnamefont {M.~J.}\ \bibnamefont
  {Shelley}}, \bibinfo {author} {\bibfnamefont {F.-R.}\ \bibnamefont {Tian}},\
  and\ \bibinfo {author} {\bibfnamefont {K.}~\bibnamefont {Wlodarski}},\
  }\bibfield  {title} {\bibinfo {title} {{Hele-Shaw flow and pattern formation
  in a time-dependent gap}},\ }\href
  {https://doi.org/10.1088/0951-7715/10/6/005} {\bibfield  {journal} {\bibinfo
  {journal} {Nonlinearity}\ }\textbf {\bibinfo {volume} {10}},\ \bibinfo
  {pages} {1471} (\bibinfo {year} {1997})}\BibitemShut {NoStop}%
\bibitem [{\citenamefont {Anjos}\ \emph {et~al.}(2022)\citenamefont {Anjos},
  \citenamefont {Zhao}, \citenamefont {Lowengrub},\ and\ \citenamefont
  {Li}}]{azll22}%
  \BibitemOpen
  \bibfield  {author} {\bibinfo {author} {\bibfnamefont {P.~H.~A.}\
  \bibnamefont {Anjos}}, \bibinfo {author} {\bibfnamefont {M.}~\bibnamefont
  {Zhao}}, \bibinfo {author} {\bibfnamefont {J.}~\bibnamefont {Lowengrub}},\
  and\ \bibinfo {author} {\bibfnamefont {S.}~\bibnamefont {Li}},\ }\bibfield
  {title} {\bibinfo {title} {Electrically controlled self-similar evolution of
  viscous fingering patterns},\ }\href
  {https://doi.org/10.1103/PhysRevFluids.7.053903} {\bibfield  {journal}
  {\bibinfo  {journal} {Phys. Rev. Fluids}\ }\textbf {\bibinfo {volume} {7}},\
  \bibinfo {pages} {053903} (\bibinfo {year} {2022})}\BibitemShut {NoStop}%
\bibitem [{\citenamefont {Jackson}\ \emph {et~al.}(1994)\citenamefont
  {Jackson}, \citenamefont {Goldstein},\ and\ \citenamefont {Cebers}}]{JGC94}%
  \BibitemOpen
  \bibfield  {author} {\bibinfo {author} {\bibfnamefont {D.~P.}\ \bibnamefont
  {Jackson}}, \bibinfo {author} {\bibfnamefont {R.~E.}\ \bibnamefont
  {Goldstein}},\ and\ \bibinfo {author} {\bibfnamefont {A.~O.}\ \bibnamefont
  {Cebers}},\ }\bibfield  {title} {\bibinfo {title} {Hydrodynamics of fingering
  instabilities in dipolar fluids},\ }\href
  {https://doi.org/10.1103/PhysRevE.50.298} {\bibfield  {journal} {\bibinfo
  {journal} {Phys. Rev. E}\ }\textbf {\bibinfo {volume} {50}},\ \bibinfo
  {pages} {298} (\bibinfo {year} {1994})}\BibitemShut {NoStop}%
\bibitem [{\citenamefont {Knobloch}\ and\ \citenamefont
  {Krechetnikov}(2015)}]{KK15}%
  \BibitemOpen
  \bibfield  {author} {\bibinfo {author} {\bibfnamefont {E.}~\bibnamefont
  {Knobloch}}\ and\ \bibinfo {author} {\bibfnamefont {R.}~\bibnamefont
  {Krechetnikov}},\ }\bibfield  {title} {\bibinfo {title} {Problems on
  time-varying domains: Formulation, dynamics, and challenges},\ }\href
  {https://doi.org/10.1007/s10440-014-9993-x} {\bibfield  {journal} {\bibinfo
  {journal} {Acta Appl. Math.}\ }\textbf {\bibinfo {volume} {137}},\ \bibinfo
  {pages} {123} (\bibinfo {year} {2015})}\BibitemShut {NoStop}%
\bibitem [{\citenamefont {Ghadiri}\ and\ \citenamefont
  {Krechetnikov}(2019)}]{GK19}%
  \BibitemOpen
  \bibfield  {author} {\bibinfo {author} {\bibfnamefont {M.}~\bibnamefont
  {Ghadiri}}\ and\ \bibinfo {author} {\bibfnamefont {R.}~\bibnamefont
  {Krechetnikov}},\ }\bibfield  {title} {\bibinfo {title} {Pattern formation on
  time-dependent domains},\ }\href {https://doi.org/10.1017/jfm.2019.659}
  {\bibfield  {journal} {\bibinfo  {journal} {J. Fluid Mech.}\ }\textbf
  {\bibinfo {volume} {880}},\ \bibinfo {pages} {136–179} (\bibinfo {year}
  {2019})}\BibitemShut {NoStop}%
\bibitem [{\citenamefont {Knobloch}\ and\ \citenamefont
  {Krechetnikov}(2014)}]{KK2014}%
  \BibitemOpen
  \bibfield  {author} {\bibinfo {author} {\bibfnamefont {E.}~\bibnamefont
  {Knobloch}}\ and\ \bibinfo {author} {\bibfnamefont {R.}~\bibnamefont
  {Krechetnikov}},\ }\bibfield  {title} {\bibinfo {title} {Stability on
  time-dependent domains},\ }\href {https://doi.org/10.1007/s00332-014-9197-6}
  {\bibfield  {journal} {\bibinfo  {journal} {J. Nonlinear Sci.}\ }\textbf
  {\bibinfo {volume} {24}},\ \bibinfo {pages} {493} (\bibinfo {year}
  {2014})}\BibitemShut {NoStop}%
\bibitem [{\citenamefont {Lobry}(1991)}]{L91}%
  \BibitemOpen
  \bibfield  {author} {\bibinfo {author} {\bibfnamefont {C.}~\bibnamefont
  {Lobry}},\ }\bibfield  {title} {\bibinfo {title} {Dynamic bifurcations},\
  }in\ \href {https://doi.org/10.1007/BFb0085020} {\emph {\bibinfo {booktitle}
  {Dynamic Bifurcations: Proceedings of a Conference held in Luminy, France,
  March 5-10, 1990}}},\ \bibinfo {series} {Lecture Notes in Mathematics}, Vol.\
  \bibinfo {volume} {1493},\ \bibinfo {editor} {edited by\ \bibinfo {editor}
  {\bibfnamefont {E.}~\bibnamefont {Beno{\^i}t}}}\ (\bibinfo  {publisher}
  {Springer},\ \bibinfo {address} {Berlin/Heidelberg},\ \bibinfo {year}
  {1991})\ pp.\ \bibinfo {pages} {1--13}\BibitemShut {NoStop}%
\bibitem [{\citenamefont {Lira}\ and\ \citenamefont {Miranda}(2012)}]{LM12}%
  \BibitemOpen
  \bibfield  {author} {\bibinfo {author} {\bibfnamefont {S.~A.}\ \bibnamefont
  {Lira}}\ and\ \bibinfo {author} {\bibfnamefont {J.~A.}\ \bibnamefont
  {Miranda}},\ }\bibfield  {title} {\bibinfo {title} {{Nonlinear traveling
  waves in confined ferrofluids}},\ }\href
  {https://doi.org/10.1103/PhysRevE.86.056301} {\bibfield  {journal} {\bibinfo
  {journal} {Phys. Rev. E}\ }\textbf {\bibinfo {volume} {86}},\ \bibinfo
  {pages} {056301} (\bibinfo {year} {2012})}\BibitemShut {NoStop}%
\bibitem [{\citenamefont {Qiu}\ \emph {et~al.}(2018)\citenamefont {Qiu},
  \citenamefont {Afkhami}, \citenamefont {Chen},\ and\ \citenamefont
  {Feng}}]{QACF18}%
  \BibitemOpen
  \bibfield  {author} {\bibinfo {author} {\bibfnamefont {M.}~\bibnamefont
  {Qiu}}, \bibinfo {author} {\bibfnamefont {S.}~\bibnamefont {Afkhami}},
  \bibinfo {author} {\bibfnamefont {C.-Y.}\ \bibnamefont {Chen}},\ and\
  \bibinfo {author} {\bibfnamefont {J.~J.}\ \bibnamefont {Feng}},\ }\bibfield
  {title} {\bibinfo {title} {Interaction of a pair of ferrofluid drops in a
  rotating magnetic field},\ }\href {https://doi.org/10.1017/jfm.2018.261}
  {\bibfield  {journal} {\bibinfo  {journal} {J. Fluid Mech.}\ }\textbf
  {\bibinfo {volume} {846}},\ \bibinfo {pages} {121} (\bibinfo {year}
  {2018})}\BibitemShut {NoStop}%
\bibitem [{\citenamefont {Roberts}(1983)}]{R83}%
  \BibitemOpen
  \bibfield  {author} {\bibinfo {author} {\bibfnamefont {A.~J.}\ \bibnamefont
  {Roberts}},\ }\bibfield  {title} {\bibinfo {title} {{A stable and accurate
  numerical method to calculate the motion of a sharp interface between
  fluids}},\ }\href {https://doi.org/10.1093/imamat/31.1.13} {\bibfield
  {journal} {\bibinfo  {journal} {IMA J. Appl. Math.}\ }\textbf {\bibinfo
  {volume} {31}},\ \bibinfo {pages} {13} (\bibinfo {year} {1983})}\BibitemShut
  {NoStop}%
\bibitem [{\citenamefont {Tryggvason}\ and\ \citenamefont {Aref}(1983)}]{TA83}%
  \BibitemOpen
  \bibfield  {author} {\bibinfo {author} {\bibfnamefont {G.}~\bibnamefont
  {Tryggvason}}\ and\ \bibinfo {author} {\bibfnamefont {H.}~\bibnamefont
  {Aref}},\ }\bibfield  {title} {\bibinfo {title} {{Numerical experiments on
  Hele Shaw flow with a sharp interface}},\ }\href
  {https://doi.org/10.1017/S0022112083002037} {\bibfield  {journal} {\bibinfo
  {journal} {J. Fluid Mech.}\ }\textbf {\bibinfo {volume} {136}},\ \bibinfo
  {pages} {1} (\bibinfo {year} {1983})}\BibitemShut {NoStop}%
\bibitem [{\citenamefont {Prosperetti}(2002)}]{Prosp02}%
  \BibitemOpen
  \bibfield  {author} {\bibinfo {author} {\bibfnamefont {A.}~\bibnamefont
  {Prosperetti}},\ }\bibfield  {title} {\bibinfo {title} {{Boundary Integral
  Methods}},\ }in\ \href {https://doi.org/10.1007/978-3-7091-2594-6{\_}7}
  {\emph {\bibinfo {booktitle} {Drop-Surface Interactions}}},\ \bibinfo
  {series} {CISM International Centre for Mechanical Sciences}, Vol.\ \bibinfo
  {volume} {456},\ \bibinfo {editor} {edited by\ \bibinfo {editor}
  {\bibfnamefont {M.}~\bibnamefont {Rein}}}\ (\bibinfo  {publisher}
  {Springer-Verlag},\ \bibinfo {address} {Wien},\ \bibinfo {year} {2002})\ pp.\
  \bibinfo {pages} {219--235}\BibitemShut {NoStop}%
\bibitem [{\citenamefont {Virtanen}\ \emph {et~al.}(2020)\citenamefont
  {Virtanen}, \citenamefont {Gommers}, \citenamefont {Oliphant}, \citenamefont
  {Haberland}, \citenamefont {Reddy}, \citenamefont {Cournapeau}, \citenamefont
  {Burovski}, \citenamefont {Peterson}, \citenamefont {Weckesser},
  \citenamefont {Bright}, \citenamefont {van~der Walt}, \citenamefont {Brett},
  \citenamefont {Wilson}, \citenamefont {Millman}, \citenamefont {Mayorov},
  \citenamefont {Nelson}, \citenamefont {Jones}, \citenamefont {Kern},
  \citenamefont {Larson}, \citenamefont {Carey}, \citenamefont {Polat},
  \citenamefont {Feng}, \citenamefont {Moore}, \citenamefont {VanderPlas},
  \citenamefont {Laxalde}, \citenamefont {Perktold}, \citenamefont {Cimrman},
  \citenamefont {Henriksen}, \citenamefont {Quintero}, \citenamefont {Harris},
  \citenamefont {Archibald}, \citenamefont {Ribeiro}, \citenamefont
  {Pedregosa},\ and\ \citenamefont {van Mulbregt}}]{SciPy}%
  \BibitemOpen
  \bibfield  {author} {\bibinfo {author} {\bibfnamefont {P.}~\bibnamefont
  {Virtanen}}, \bibinfo {author} {\bibfnamefont {R.}~\bibnamefont {Gommers}},
  \bibinfo {author} {\bibfnamefont {T.~E.}\ \bibnamefont {Oliphant}}, \bibinfo
  {author} {\bibfnamefont {M.}~\bibnamefont {Haberland}}, \bibinfo {author}
  {\bibfnamefont {T.}~\bibnamefont {Reddy}}, \bibinfo {author} {\bibfnamefont
  {D.}~\bibnamefont {Cournapeau}}, \bibinfo {author} {\bibfnamefont
  {E.}~\bibnamefont {Burovski}}, \bibinfo {author} {\bibfnamefont
  {P.}~\bibnamefont {Peterson}}, \bibinfo {author} {\bibfnamefont
  {W.}~\bibnamefont {Weckesser}}, \bibinfo {author} {\bibfnamefont
  {J.}~\bibnamefont {Bright}}, \bibinfo {author} {\bibfnamefont {S.~J.}\
  \bibnamefont {van~der Walt}}, \bibinfo {author} {\bibfnamefont
  {M.}~\bibnamefont {Brett}}, \bibinfo {author} {\bibfnamefont
  {J.}~\bibnamefont {Wilson}}, \bibinfo {author} {\bibfnamefont {K.~J.}\
  \bibnamefont {Millman}}, \bibinfo {author} {\bibfnamefont {N.}~\bibnamefont
  {Mayorov}}, \bibinfo {author} {\bibfnamefont {A.~R.~J.}\ \bibnamefont
  {Nelson}}, \bibinfo {author} {\bibfnamefont {E.}~\bibnamefont {Jones}},
  \bibinfo {author} {\bibfnamefont {R.}~\bibnamefont {Kern}}, \bibinfo {author}
  {\bibfnamefont {E.}~\bibnamefont {Larson}}, \bibinfo {author} {\bibfnamefont
  {C.~J.}\ \bibnamefont {Carey}}, \bibinfo {author} {\bibfnamefont
  {I.}~\bibnamefont {Polat}}, \bibinfo {author} {\bibfnamefont
  {Y.}~\bibnamefont {Feng}}, \bibinfo {author} {\bibfnamefont {E.~W.}\
  \bibnamefont {Moore}}, \bibinfo {author} {\bibfnamefont {J.}~\bibnamefont
  {VanderPlas}}, \bibinfo {author} {\bibfnamefont {D.}~\bibnamefont {Laxalde}},
  \bibinfo {author} {\bibfnamefont {J.}~\bibnamefont {Perktold}}, \bibinfo
  {author} {\bibfnamefont {R.}~\bibnamefont {Cimrman}}, \bibinfo {author}
  {\bibfnamefont {I.}~\bibnamefont {Henriksen}}, \bibinfo {author}
  {\bibfnamefont {E.~A.}\ \bibnamefont {Quintero}}, \bibinfo {author}
  {\bibfnamefont {C.~R.}\ \bibnamefont {Harris}}, \bibinfo {author}
  {\bibfnamefont {A.~M.}\ \bibnamefont {Archibald}}, \bibinfo {author}
  {\bibfnamefont {A.~H.}\ \bibnamefont {Ribeiro}}, \bibinfo {author}
  {\bibfnamefont {F.}~\bibnamefont {Pedregosa}},\ and\ \bibinfo {author}
  {\bibfnamefont {P.}~\bibnamefont {van Mulbregt}},\ }\bibfield  {title}
  {\bibinfo {title} {{SciPy 1.0: fundamental algorithms for scientific
  computing in Python}},\ }\href {https://doi.org/10.1038/s41592-019-0686-2}
  {\bibfield  {journal} {\bibinfo  {journal} {Nat. Methods}\ }\textbf {\bibinfo
  {volume} {17}},\ \bibinfo {pages} {261} (\bibinfo {year} {2020})}\BibitemShut
  {NoStop}%
\bibitem [{\citenamefont {Kuznetsov}(1998)}]{KKK98}%
  \BibitemOpen
  \bibfield  {author} {\bibinfo {author} {\bibfnamefont {Y.~A.}\ \bibnamefont
  {Kuznetsov}},\ }\href {https://doi.org/10.1007/978-1-4757-3978-7} {\emph
  {\bibinfo {title} {Elements of Applied Bifurcation Theory}}},\ \bibinfo
  {edition} {3rd}\ ed.,\ \bibinfo {series} {Applied Mathematical Sciences},
  Vol.\ \bibinfo {volume} {112}\ (\bibinfo  {publisher} {Springer},\ \bibinfo
  {address} {New York, NY},\ \bibinfo {year} {1998})\BibitemShut {NoStop}%
\bibitem [{\citenamefont {Wiggins}(2003)}]{WWG03}%
  \BibitemOpen
  \bibfield  {author} {\bibinfo {author} {\bibfnamefont {S.}~\bibnamefont
  {Wiggins}},\ }\href {https://doi.org/10.1007/b97481} {\emph {\bibinfo {title}
  {Introduction to Applied Nonlinear Dynamical Systems and Chaos}}},\ \bibinfo
  {series} {Texts in Applied Mathematics}, Vol.~\bibinfo {volume} {2}\
  (\bibinfo  {publisher} {Springer},\ \bibinfo {address} {New York, NY},\
  \bibinfo {year} {2003})\BibitemShut {NoStop}%
\bibitem [{\citenamefont {\'{A}lvarez Lacalle}\ \emph
  {et~al.}(2004)\citenamefont {\'{A}lvarez Lacalle}, \citenamefont
  {Ort\'{i}n},\ and\ \citenamefont {Casademunt}}]{AOC04}%
  \BibitemOpen
  \bibfield  {author} {\bibinfo {author} {\bibfnamefont {E.}~\bibnamefont
  {\'{A}lvarez Lacalle}}, \bibinfo {author} {\bibfnamefont {J.}~\bibnamefont
  {Ort\'{i}n}},\ and\ \bibinfo {author} {\bibfnamefont {J.}~\bibnamefont
  {Casademunt}},\ }\bibfield  {title} {\bibinfo {title} {Nonlinear
  {Saffman}-{Taylor} instability},\ }\href
  {https://doi.org/10.1103/PhysRevLett.92.054501} {\bibfield  {journal}
  {\bibinfo  {journal} {Phys. Rev. Lett.}\ }\textbf {\bibinfo {volume} {92}},\
  \bibinfo {pages} {054501} (\bibinfo {year} {2004})}\BibitemShut {NoStop}%
\bibitem [{\citenamefont {Baer}\ \emph {et~al.}(1989)\citenamefont {Baer},
  \citenamefont {Erneux},\ and\ \citenamefont {Rinzel}}]{BER89}%
  \BibitemOpen
  \bibfield  {author} {\bibinfo {author} {\bibfnamefont {S.~M.}\ \bibnamefont
  {Baer}}, \bibinfo {author} {\bibfnamefont {T.}~\bibnamefont {Erneux}},\ and\
  \bibinfo {author} {\bibfnamefont {J.}~\bibnamefont {Rinzel}},\ }\bibfield
  {title} {\bibinfo {title} {The slow passage through a {Hopf} bifurcation:
  delay, memory effects, and resonance},\ }\href
  {https://doi.org/10.1137/0149003} {\bibfield  {journal} {\bibinfo  {journal}
  {SIAM J. Appl. Math.}\ }\textbf {\bibinfo {volume} {49}},\ \bibinfo {pages}
  {55} (\bibinfo {year} {1989})}\BibitemShut {NoStop}%
\bibitem [{\citenamefont {Kevorkian}\ and\ \citenamefont {Cole}(1996)}]{KC96}%
  \BibitemOpen
  \bibfield  {author} {\bibinfo {author} {\bibfnamefont {J.}~\bibnamefont
  {Kevorkian}}\ and\ \bibinfo {author} {\bibfnamefont {J.~D.}\ \bibnamefont
  {Cole}},\ }\href {https://doi.org/10.1007/978-1-4612-3968-0} {\emph {\bibinfo
  {title} {{Multiple Scale and Singular Perturbation Methods}}}},\ \bibinfo
  {series} {Applied Mathematical Sciences}, Vol.\ \bibinfo {volume} {114}\
  (\bibinfo  {publisher} {Springer New York},\ \bibinfo {address} {New York,
  NY},\ \bibinfo {year} {1996})\BibitemShut {NoStop}%
\bibitem [{\citenamefont {Landau}(1944)}]{landau1944}%
  \BibitemOpen
  \bibfield  {author} {\bibinfo {author} {\bibfnamefont {L.~D.}\ \bibnamefont
  {Landau}},\ }\bibfield  {title} {\bibinfo {title} {On the problem of
  turbulence},\ }\href@noop {} {\bibfield  {journal} {\bibinfo  {journal}
  {Dokl. Akad. Nauk SSSR}\ }\textbf {\bibinfo {volume} {44}},\ \bibinfo {pages}
  {339} (\bibinfo {year} {1944})},\ \bibinfo {note} {in Russian}\BibitemShut
  {NoStop}%
\bibitem [{\citenamefont {Weisstein}(2022)}]{erfi}%
  \BibitemOpen
  \bibfield  {author} {\bibinfo {author} {\bibfnamefont {E.~W.}\ \bibnamefont
  {Weisstein}},\ }\href {https://mathworld.wolfram.com/Erfi.html} {\bibinfo
  {title} {{Erfi}}},\ \bibinfo {howpublished} {MathWorld--A Wolfram Web
  Resource} (\bibinfo {year} {2022})\BibitemShut {NoStop}%
\bibitem [{\citenamefont {Taylor}(1967)}]{Taylor_NCFMF}%
  \BibitemOpen
  \bibfield  {author} {\bibinfo {author} {\bibfnamefont {G.~I.}\ \bibnamefont
  {Taylor}},\ }\href {http://web.mit.edu/hml/ncfmf.html} {\bibinfo {title}
  {{Low {Reynolds} Number Flows}}},\ \bibinfo {howpublished} {National
  Committee for Fluid Mechanics Films, Education Development Center, Inc.}
  (\bibinfo {year} {1967})\BibitemShut {NoStop}%
\bibitem [{\citenamefont {Bacri}\ and\ \citenamefont {Salin}(1984)}]{BS84}%
  \BibitemOpen
  \bibfield  {author} {\bibinfo {author} {\bibfnamefont {J.-C.}\ \bibnamefont
  {Bacri}}\ and\ \bibinfo {author} {\bibfnamefont {D.}~\bibnamefont {Salin}},\
  }\bibfield  {title} {\bibinfo {title} {First-order transition in the
  instability of a magnetic fluid interface},\ }\href
  {https://doi.org/10.1051/jphyslet:019840045011055900} {\bibfield  {journal}
  {\bibinfo  {journal} {J. Physique Lett.}\ }\textbf {\bibinfo {volume} {45}},\
  \bibinfo {pages} {559} (\bibinfo {year} {1984})}\BibitemShut {NoStop}%
\end{thebibliography}%

\balancecolsandclearpage
\appendix
\section{Coefficients for the reduced model}
\label{sec:ai bi ci di}
The coefficients in the system of ODEs \eqref{eq:4th order ode} are
\begin{equation*}\begin{aligned}
a_1&=\Lambda(k),\\
a_2&=F(k,-k ) + F(k,2k) \\
&\qquad\qquad+ G(k,-k ) \Lambda(-k )   + G(k,2k)\Lambda(2k),\\
a_3&=F(k,-2k) + F(k,3k) \\
&\qquad\qquad+ G(k,-2k) \Lambda(-2k)  + G(k,3k)\Lambda(3k),\\
a_4&=F(k,-3k) + F(k,4k) \\
&\qquad\qquad+ G(k,-3k) \Lambda(-3k)   + G(k,4k)\Lambda(4k),\\
b_1&=\Lambda(2k), \\
b_2&=F(2k,-k )	+	F(2k,3k)	\\
&\qquad\qquad+ G(2k,-k)	\Lambda(-k)  + G(2k,3k)\Lambda(3k),\\
b_3&=F(2k,-2k)	+	F(2k,4k)\\	
&\qquad\qquad+  G(2k,-2k)	\Lambda(-2k) + G(2k,4k)\Lambda(4k),\\
b_4&=F(2k,k 	)	+	G(2k,k)	\Lambda(k),\\
c_1&=\Lambda(3k), \\
c_2&=F(3k,-k)	+F(3k,4k)  \\
&\qquad\qquad+  G(3k,-k) \Lambda(-k)  + G(3k,4k)\Lambda(4k),\\
c_3&=F(3k,k )	+F(3k,2k) \\
&\qquad\qquad+ G(3k,k ) \Lambda(k)   + G(3k,2k)\Lambda(2k), \\
d_1&=\Lambda(4k), \\
d_2&=F(4k,k )	+F(4k,3k)\\
&\qquad\qquad+ G(4k,k )\Lambda(k) 	+ G(4k,3k)\Lambda(3k),\\
d_3&=F(4k,2k)	+ G(4k,2k)\Lambda(2k),
\end{aligned}\end{equation*}
where the functions $F$ and $G$ are given in \cite{YC21}.  

\section{Four-mode equation of motion in polar coordinate}
\label{sec:polar coordinate}
The four-mode coupled system \eqref{eq:4th order ode} written in polar is
\begin{subequations}\label{eq:4th order ode_polar}\begin{align}
\dot{r}_x+i\dot{\phi}_xr_x
&=a_1r_x +a_2r_xr_ye^{i(\phi_y-2\phi_x)}\\
&\phantom{=}+a_3r_yr_ze^{i(\phi_z-\phi_y-\phi_x)}
+a_4r_zr_pe^{i(\phi_p-\phi_z-\phi_x)},\nonumber\\
\dot{r}_y+i\dot{\phi}_yr_y
&=b_1r_y +b_2r_xr_ze^{i(\phi_z-\phi_x-\phi_y)}\\
&\phantom{=}+b_3r_yr_pe^{i(\phi_p-2\phi_y)}
+b_4r_x^2e^{i(2\phi_x-\phi_y)},\nonumber\\
\dot{r}_z+i\dot{\phi}_zr_z
&=c_1r_z +c_2r_xr_pe^{i(\phi_p-\phi_x-\phi_z)}\\
&\phantom{=}+c_3r_xr_ye^{i(\phi_x+\phi_y-\phi_z)},\nonumber\\
\dot{r}_p+i\dot{\phi}_pr_p
&=d_1r_p +d_2r_xr_ze^{i(\phi_x+\phi_z-\phi_p)}\\
&\phantom{=}+d_3r_y^2e^{i(2\phi_y-\phi_p)}.\nonumber
\end{align}\end{subequations}%

\section{Eigenvalues of perturbation growth matrix}
\label{sec:eigenvalues}
\begin{figure}[h]
    \centering
    \includegraphics[keepaspectratio=true,width=\columnwidth]{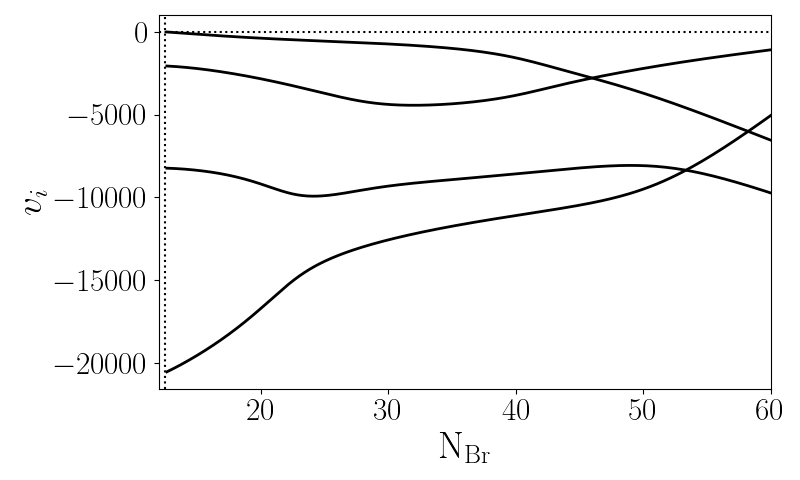}
    \caption{The real part of the four eigenvalues of traveling wave solution stability matrix $\bm M$ given in Eq.~\eqref{eq:matrix M}.} 
    \label{fig:eigenvalue}
\end{figure}

The matrix $\bm{M}$ governing the evolution of perturbations,  $\dot{\bm{\epsilon}}=\bm{M}\bm{\epsilon}$, in Eq.~\eqref{eq:perturbed traveling wave} is
\begin{equation}
\bm{M}
=
\begin{bsmallmatrix}
    a_1-i\Omega +a_2R_y & a_2R_x+a_3R_z &
    a_3R_y+a_4Rp & a_4 R_z\\
    b_2R_z +2b_4R_x & b_1-2i\Omega +b_3R_p &
    b_2R_x& b_3R_y\\
    c_2R_p+c_3R_y & c_3 R_x & (c_1-3i\Omega)& c_2R_x\\
    d_2R_z & 2d_3R_y & d_2R_x & d_1-4i\Omega
\end{bsmallmatrix}.
\label{eq:matrix M}
\end{equation}
The real part of the four eigenvalues, $\{v_i=\Re[\eig(\bm{M})]\}_{i=1,2,3,4}$, are plotted in Fig.~\ref{fig:eigenvalue} as functions of $\nbr\in [12.5,60]$.

\section{Center manifold derivation}
\label{sec:Center manifold derivation}
Assume the dynamics on the center manifold can be related by a scalar function $y=V(x,x^*)$. To quadratic order, its Taylor series is
\begin{equation}
    V(x,x^*) =\frac{1}{2}g_{20}x^2
    +g_{11} xx^* 
    +\frac{1}{2}g_{02}x^{*2}+\mathcal{O}(|x|^3).
    \label{eq:v(x,x*)}
\end{equation}
The unknown coefficient $g_{20}$, $g_{11}$, and $g_{02}$ can be found by substituting \eqref{eq:v(x,x*)} into the reduced system \eqref{eq:2nd order ode}:
\begin{subequations}\begin{align}
    \dot{y} &= V_x \dot{x} +V_{x^*} \dot{x}^*\\
    &= a_1 g_{20}x^2 +2\Re(a_1)g_{11}xx^*+a_1^* g_{02}x^{*2}+\mathcal{O}(|x|^3),\nonumber\\
    \dot{y} &= b_1y+b_4x^2\\
    &=
    \left(\frac{b_1}{2}g_{20}+b_4\right)x^2 +b_1g_{11}xx^* +\frac{b_1}{2}g_{20}x^{*2}+\mathcal{O}(|x|^3).\nonumber
\end{align}\label{eq:v system}\end{subequations}
The equivalence of the two equations in system~\eqref{eq:v system} at $\mathcal{O}(|x|^2)$ gives $g_{20}=2b_4/(2a_1- b_1)$ and $g_{11}=g_{02}=0$.
\begin{figure*}
    \centering
    \includegraphics[keepaspectratio=true,width=\textwidth]{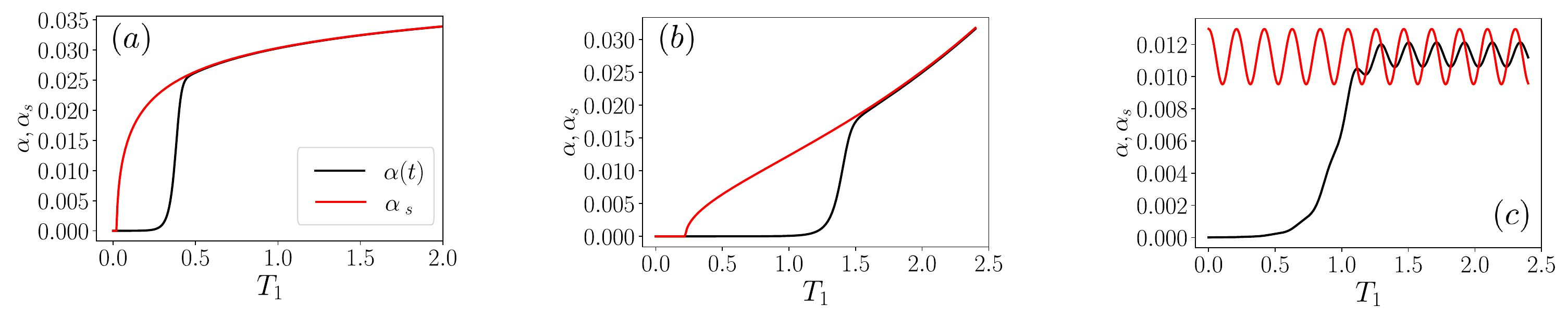}
    \caption{The time-dependent solution $\alpha(T_1)$ (black) and quasi-static solution $\alpha_s$ (red) evaluated from Eq.~\eqref{eq:Landau} with (a) $\varkappa=\varkappa_0 +20\log (20T_1+1)$; (b)$\varkappa=\varkappa_0 +0.08 I e^{T_1}$; and (c) $\varkappa=\varkappa_0 +0.03I\cos(30T_1)$. $I=75$ for all three cases, and $\varkappa_0=7.5$ for (a,b), while $\kappa_0=-7.5$ for (c). } 
    \label{fig:protocols}
\end{figure*}
\section{Amplitude equation via multiple-time-scale analsys}
\label{sec:multi-scale}
Substituting the expansion~\eqref{eq:expansion} into the small amplitude equations~\eqref{eq:small amplitude}, we obtain the system
\begin{subequations}\begin{align}
  &\left(\pp{}{t}+\epsilon^2\pp{}{T_1}\right)
 (x_0+\epsilon x_1 +\epsilon^2x_2)\\
 &\quad=(\varkappa\epsilon^2+i\omega)(x_0+\epsilon x_1+\epsilon^2x_2)\nonumber\\
 &\qquad+\epsilon a_2 (x_0^*+\epsilon x_1^*+\epsilon^2x_2^*)
  (y_0+\epsilon y_1 +\epsilon^2y_2),\nonumber\\
   &\left(\pp{}{t}+\epsilon^2\pp{}{T_1}\right)
 \left(y_0+\epsilon y_1 +\epsilon^2y_2\right)\\
 &\quad=b_1(y_0+\epsilon y_1+\epsilon^2y_2)+
 \epsilon b_4 
 (x_0+\epsilon x_1 +\epsilon^2x_2)^2.\nonumber
\end{align}\end{subequations}
By collecting terms at $\mathcal{O}(1)$, we obtain the leading-order equation~\eqref{eq:O(0) eq} and its solution~\eqref{eq:O(0) sol}.

Then, at $\mathcal{O}(\epsilon)$, the equation is
\begin{subequations}\begin{align}
 \pp{x_1}{t}-i\omega x_1 &= a_2 x_0^*y_0
    =a_2 A_x^*A_y e^{(b_1-i\omega)t},\\
 \pp{y_1}{t}-b_1 y_1 &= b_4x_0^2 = b_4A_x^2 e^{2i\omega t},
\end{align}\end{subequations}
which can be solved as
\begin{subequations}\begin{align}
    x_1&=\frac{a_2}{b_1-2i\omega}
    \left(A_x^*A_ye^{(b_1-i\omega)t}
    -X^*Y e^{i\omega t}\right),\\
    y_1&=\frac{-b_4}{b_1-2i\omega}
    \left(A_x^2e^{2i\omega t}-X^2e^{b_1 t}\right),
\end{align}\end{subequations}
with initial condition $x_1(0,0)=0$, $y_1(0,0)=0$.

Finally, at $\mathcal{O}(\epsilon^2)$, we have
\begin{subequations}\begin{align}
\pp{x_2}{t}-i\omega x_2&=-\pp{x_0}{T_1}+\varkappa x_0+
 a_2(x_0^* y_1+ x_1^* y_0),\label{eq:x2}
 \\
  \pp{y_2}{t}-b_1 y_2&=-\pp{}{T_1}y_0 +2b_4 x_0x_1,\label{eq:O(2)y}
\end{align}\end{subequations}
with initial condition $x_2(0,0)=0$, $y_2(0,0)=0$.
The nonlinear term in Eq.~\eqref{eq:x2} can be calculated as:
\begin{eqnarray}
 &\;&x_0^* y_1+ x_1^* y_0\\
 &=&
 A_x^*e^{-i\omega t} \left[\frac{-b_4}{b_1-2i\omega}
   (A_x^2e^{2i\omega t}-X^2e^{b_1 t})\right]\nonumber\\
   &+&A_y(T_1)e^{b_1 t}
   \left[\frac{a_2^*}{b_1^*+2i\omega}
    (A_xA_y^*e^{(b_1^*+i\omega)t}
    -XY^* e^{-i\omega t})\right].\nonumber
\end{eqnarray}
To eliminate the secular term, we require that
\begin{equation}
    -\pp{x_0}{T_1} + \varkappa x_0 + \frac{a_2b_4}{2i\omega-b_1}A_x^* A_x^2 e^{i\omega t} = 0,
\end{equation}
which yields the amplitude equation
\begin{equation}
    \frac{d A_x}{dT_1}  = \varkappa A_x + \frac{a_2b_4}{2i\omega-b_1} |A_x|^2 A_x.
    \label{eq:D1 Ax}
\end{equation}

Letting $A_x = \alpha e^{i\beta}$, we have%
\begin{equation}
    \frac{dA_x}{dT_1} = \left[ \frac{d \alpha}{dT_1} + i \frac{d\beta}{dT_1} \alpha \right] e^{i\beta}, 
\end{equation}
and Eq.~\eqref{eq:D1 Ax} becomes%
\begin{subequations}\begin{align}
  \frac{d\alpha}{dT_1}  &= \varkappa \alpha + \Re\left(\frac{a_2b_4}{2i\omega-b_1}\right)\alpha^3,\\
  \frac{d\beta}{dT_1}  &= \Im\left(\frac{a_2b_4}{2i\omega-b_1}\right)\alpha^2.
\end{align}\end{subequations}

\section{Other possible time-varying protocols}
\label{sec:protocols}
{The analysis can be carried out for arbitrary time-varying protocols. In this appendix, we show three examples: a log-varying, an exponentially increasing, and an oscillating growth rate.
We can observe that for a log-varying or an exponentially increasing growth rate, as in Fig.~\ref{fig:protocols}(a) and (b), respectively, $\alpha$ will saturate to the quasi-static solution $\alpha_s$. This is not the case, however, for the oscillating growth rate shown in Fig.~\ref{fig:protocols}(c).
This observation opens a series of follow-up questions: (i) How do we prove the saturation mathematically, and how do we obtain the explicit delay prediction like in Eq.~\eqref{eq:T_1^e simple}? (ii) How do we quantify the reliable prediction time range (since the exponential variation will quickly break down the slow-time-variation assumption)? (iii) How do we quantify the observed phase lag between the time-dependent solution and the quasi-static solution for an oscillating growth rate? These questions are left to future work.}

\end{document}